\documentclass[%
 reprint,
 amsmath,amssymb,
 aps,
prb,
superscriptaddress,
longbibliography
]{revtex4-1}

\usepackage{graphicx}
\usepackage{dcolumn}
\usepackage{bm}
\usepackage{bbold}
\usepackage{braket}
\usepackage[ruled,vlined]{algorithm2e} 
\usepackage[thinc]{esdiff}

\usepackage{hyperref}
\hypersetup{colorlinks, linkcolor={blue},citecolor={blue}, urlcolor={blue}}

\newcommand{\id}{\mathbb{1}} 

\begin{document}

\preprint{APS/123-QED}

\title{Matrix approach for optimal spatio-temporal coherent control of wave scattering}

\author{Cl{\'e}ment Ferise}
\affiliation{Univ Rennes, CNRS, IETR - UMR 6164, F-35000 Rennes, France}%

\author{Philipp del Hougne}
\affiliation{Univ Rennes, CNRS, IETR - UMR 6164, F-35000 Rennes, France}%

\author{Matthieu Davy}
\affiliation{Univ Rennes, CNRS, IETR - UMR 6164, F-35000 Rennes, France}%
\email{matthieu.davy@univ-rennes1.fr}

\date{\today}

\begin{abstract} 
We present and experimentally verify a matrix approach for determining how to optimally sculpt an input wavefront both in space and time for any desired wave-control functionality, irrespective of the complexity of the wave scattering. We leverage a singular value decomposition of the transport matrix that fully captures how both the spatial and temporal degrees of freedom available to shape the input wavefront impact the output wavefront's spatial and temporal form. In our experiments in the microwave domain, we use our formalism to successfully tackle three iconic wave-control tasks in a disordered cavity: (i) reflectionless transient excitation (``virtual perfect absorption''), (ii) optimal energy deposition,  and (iii) scattering-invariant time-varying states.

\end{abstract}

\maketitle

\section{Introduction}

The ability to sculpt the wavefront incident on a scattering system allows for a judicious control of the outgoing wavefront's spatial and/or temporal shape, thanks to the linearity of the wave equation~\cite{rotter2017light,Cao2022}. The crux lies in identifying the optimal input wavefront for a desired wave-control functionality. 
While iterative techniques optimize in situ the incident wavefront until the output converges to the desired one by relying on a feedback~\cite{Vellekoop2007,Vellekoop2010,Ma2014,Horstmeyer2015}, matrix-based techniques 
achieve provably optimal wave control using tools from linear algebra~\cite{Tanter2000,Vellekoop2008a,rotter2017light}. Different wave-engineering domains (acoustics, optics, etc.)~have explored matrix approaches for optimal coherent wave control under various constraints (e.g., only spatial control over the input, only aiming at focusing in space and time). Here we present a general approach based on a multi-spectral transport matrix  that can optimally leverage both spatial and temporal degrees of freedom (DoFs) on input and output sides.  

\begin{figure*}
    \centering
    \includegraphics[width=18cm]{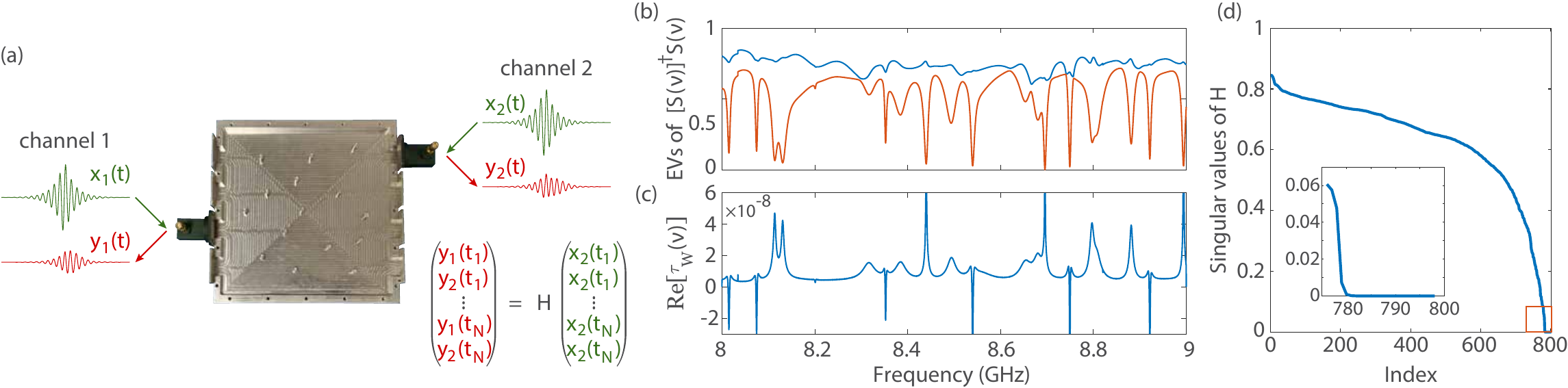}
    \caption{(a) Photographic image of our quasi-2D disordered cavity ($L=W=205.74$~mm, $h=10.16$~mm) with the top plate removed to show the inside. One coax-to-waveguide transition is attached at each side of the cavity, yielding a two-channel scattering system. The inset illustrates our spatio-temporal matrix formalism defined in Eq.~(\ref{eq:linear}). (b) Spectrum of the two eigenvalues of $\left[S(\nu)\right]^\dagger S(\nu)$. (c) Spectrum of the real part of the trace of the Wigner-Smith operator $Q(\nu)$.
    (d) Singular value spectrum of $H$ (see Eq.~(\ref{eq:linear})).}
    \label{fig:setup}
\end{figure*}

For monochromatic waves, optimal wave-control functionalities are based on measuring the scattering matrix that fully encodes the linear relation between spatial input and spatial output DoFs. 
In optics, the transmission matrix was initially measured using deformable mirrors capable of spatially shaping the input wavefront to achieve maximal focusing of light traversing a multiple scattering medium~\cite{Popoff2010}. More complicated functionalities beyond focusing are then achieved using a singular value decomposition (SVD) of the transport matrix. This includes selective focusing inside the scattering medium~\cite{popoff2011exploiting}, perfect absorption of incident radiations \cite{chong2010coherent,wan2011time}, speckle engineering~\cite{Devaud2021} and access to open and closed transmission eigenchannels~\cite{Vellekoop2008a,kim2012maximal,Gerardin2014,Sarma2016} in diffusive media, e.g., for optimal energy deposition~\cite{Jeong18,Bender2022}. New possibilities have recently emerged by applying an SVD or an eigendecomposition to other operators constructed from the transport matrix. This has led to proposals for schemes aimed at optimal energy storage~\cite{Durand2019,delHougne2021}, optimal micro-manipulation of targets embedded in scattering environments \cite{Horodynski2019NatPhot} and ultimate sensitivity of the output wavefield to perturbations \cite{Bouchet2021NatPhys} 
-- provided that the system's transport matrix can be rapidly characterized under low-noise conditions. 


Manipulating spatially the incident wavefront also enables a degree of control on output signals in the time domain for incident pulses~\cite{Aulbach2011,katz2011focusing}. Focusing and maximal energy deposition at arbitrary times are based on measuring the multispectral TM~\cite{andreoli2015deterministic,Mounaix2016a} or the time-gated TM~\cite{ChoiPRL2013,Mounaix2016,Jeong18,devaud2022temporal}. More recently, spatio-temporal control over the output has also been achieved by engineering the scattering system as opposed to sculpting the input wavefront~\cite{del2016spatiotemporal,imani2022metasurface}, but this approach is conceptually very different because the parametrization of the transport matrix via tunable scattering parameters is in general non-linear~\cite{rabault2023tacit}.

While these approaches are still limited to spatial incident DoFs, existing optimal spatio-temporal control makes use of both spatial and temporal DoFs on the input side~\cite{Mosk2012}. For the specific functionality of maximal focusing on an output channel, time reversal (TR) is an established technique pioneered in acoustics that captures the spatio-temporal input-output relation of a (usually complex) scattering system~\cite{fink1992time,Derode1995}. TR consists in emitting phase-conjugated broadband transmission spectra, and can be refined with iterative techniques~\cite{lemoult2009manipulating}. 
In reverberating media, the spatio-temporal compression of long impulse responses shows that a control on output spatial DoFs is achieved using input temporal DoFs only for focusing~\cite{Derode1995,draeger1997one,Lerosey2007} and perturbation sensing~\cite{Bouchet2022}. Corresponding results have also been achieved in the optical domain for a single-channel~\cite{McCabe2011} and multi-channel~\cite{mounaix2020time} systems. However, the TR technique does not offer any insights into how to achieve more elaborate wave control.

In this paper, we introduce an SVD-based approach enabling optimal spatio-temporal wavefront shaping for arbitrary desired wave-control functionalities. 
We illustrate the capabilities of our formalism by applying it (i) to achieve reflectionless transient excitation (``virtual perfect absorption'', VPA),  (ii) to optimally deliver energy at a targeted time, and (iii) to excite time-varying scattering-invariant states. Thereby, we generalize the VPA concept from the single-mode~\cite{Baranov2017optica} to the multi-resonance regime, as well as scattering-invariant states from the harmonic~\cite{Pai2021} to the time-varying regime. Moreover, our formalism constitutes a unifying framework to understand to-date seemingly unrelated wave-control concepts.

\section{Matrix Formalism}
For monochromatic waves with frequency $\nu$, the transport matrix $\mathcal{H}(\nu)$ describes the linear relation between the $N$ incoming channels $x(\nu)$ and $M$ outgoing channels $y(\nu)$ as $y(\nu) = \mathcal{H}(\nu) x(\nu)$. If \textit{all} connected channels are consider for both input and output, $\mathcal{H}(\nu)$ is the system's full scattering matrix $S(\nu)$. 

To extend this monochromatic input-output relation to the time domain \textit{on both input and output sides}, we consider that the $N$ time signals $x_n(t_m)$ are sampled on $N_{tx}$ points within a time interval $\Delta t_x = [0 \ t_c]$. Their spectra $x_n(\nu)$  are computed on $N_\nu$ frequencies using the discrete-Fourier-transform (DFT) operator, $x_n(\nu) = D_x x_n(t) $, where $D_{x_{km}} = e^{-i 2 \pi \nu_k t_m }/\sqrt{N_{tx}}$. 
To obtain a single multi-channel spatio-temporal linear relation, we first concatenate these vectors into the single vectors $x_t$ and $x_\nu$ of length $NN_{tx}$ and $NN_{\nu}$. The elements of $x_t$ are $x_{t_j} = x_n(t_m)$ with $j = (m-1)N + n$. The linear relation between vectors $x_t$ and $x_\nu$ is described by the sparse multi-channel DFT operator $\tilde{D}_x$ of dimensions $NN_\nu \times NN_{tx}$ (see SM), $x_\nu = \tilde{D}_x x_t$. For $N_{tx} = N_\nu$ and $t_c = t_{\mathrm{max}}=1/\delta \nu$, both $D_x$ and $\tilde{D}_x$ are unitary operators. However, in the more general case of 
$t_c < t_{\mathrm{max}}$, they are non-square matrices verifying $D_x^\dagger D_x = \id_{N_{tx}}$ and $\tilde{D}_x^\dagger \tilde{D}_x  = \id_{NN_{tx}}$ but $ D_x D_x^\dagger \neq \id_{N_\nu}$ and $\tilde{D}_x \tilde{D}_x^\dagger \neq \id_{NN_{\nu}}$. We apply the same procedure to the $M$ outgoing signals $y_n(t)$ measured within the time window $\Delta t_y$. 

Finally, we introduce the $NN_\nu \times NN_\nu$ multi-spectral block-diagonal matrix $S_D$ with the $k$th diagonal block given by $S(\nu_k)$, yielding the following linear relation,
\begin{equation}
    y_t = \tilde{D}_y^\dagger S_D \tilde{D}_x x_t = {H} x_t,
\label{eq:linear}
\end{equation}
\noindent where we have defined the operator ${H} = \tilde{D}_y^\dagger S_D \tilde{D}_x$. This formulation relates the spatial and temporal DoFs of a coherent input wavefront to the spatial and temporal DoFs of the corresponding output wavefront.  

Using Eq.~(\ref{eq:linear}), the spatio-temporal input wavefront $x$ giving any arbitrary multi-channel time-varying output signal $y_T$ can be estimated using the pseudo-inverse ${H}^+$ of ${H}$ as $x = {H}^{+} y_T$. The case of the well-established TR technique ~\cite{fink1992time} can also be interpreted in light of Eq.~(\ref{eq:linear}). In a time-reversal experiment, the goal is to maximize the signal on the $k$th output channel at time $t=t_i$ by injecting a spatio-temporally sculpted wavefront in the time interval $\Delta t_x = [0 \ t_i]$. The desired output signal is $y_n(t) = \delta(k-n) \delta(t-t_i)$. Simple algebra shows that for a pseudo-inverse ${H}^+={H}^\dagger$, the corresponding input wavefront is $x_n(t) = \tilde{D}_x S_{nk}^*$, corresponding as expected to the time-reversed impulse responses between the $N$ input ports and the $k$th output port. TR is hence a special case of our more general technique for optimal spatio-temporal wave control.

Of particular interest is the SVD of ${H}$: ${H} = U \sqrt{\Lambda} V^\dagger$. The diagonal matrix $\Lambda$ directly yields the integral of output intensities over the time interval $\Delta t_y$ and all ports
$\lambda_n = ||{H} V_n||^2 =  \int_{\Delta t_y} [\Sigma_n |y_n(t)|^2] dt$ upon injection through all ports within the time interval $\Delta t_x$ of the spatio-temporal states $x_n(t)$. The incident signals $x_n(t)$ composing the vector $V_n$ are normalized such as $\int_{\Delta t_x} [\Sigma_n |x_n(t)|^2] dt = 1$. If our transport matrix $\mathcal{H}$ is taken to be $S$, the highest and lowest values of $\lambda_n$ determine the complete range of achievable output reflection, and hence offer a simple method to identify the provably optimal input states for maximal or minimal output reflections. If instead  $\mathcal{H}$ is a transmission matrix, then $\lambda_n$ corresponds to transmitted rather than reflected intensity instead.

\begin{figure*}
    \centering
    \includegraphics[width=18cm]{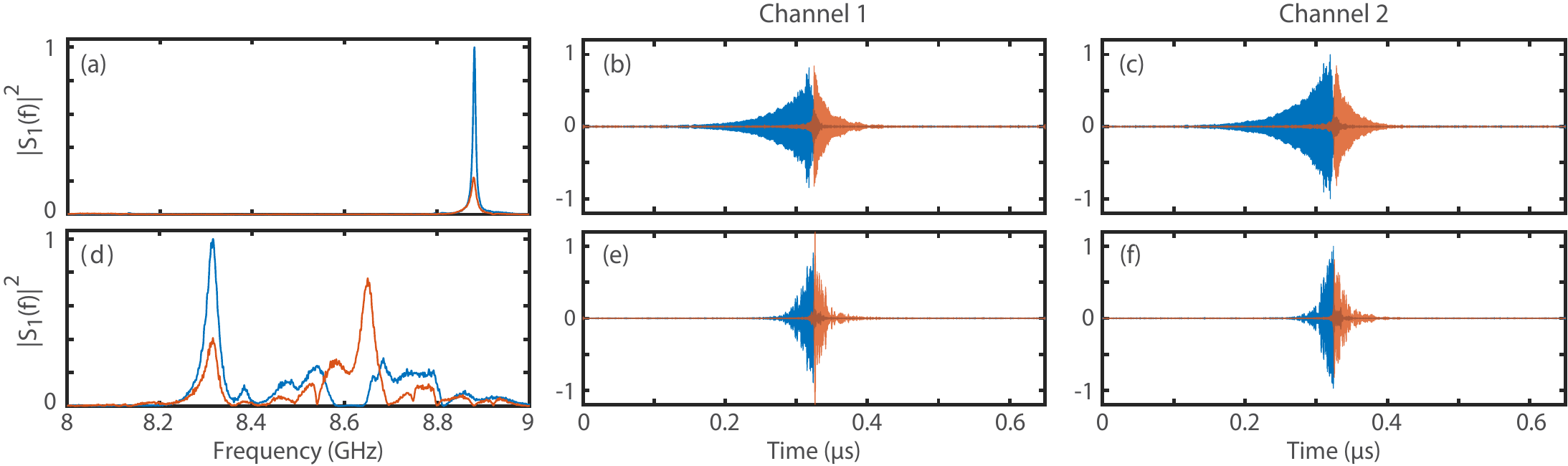}
    \caption{Experimentally measured output signals upon coherent spatio-temporal in situ injection of various right singular vectors of $H$. For each considered state, we display the injected (blue) and outcoming (red) time-varying signals on the two channels (middle and right columns), as well as the corresponding signal spectra averaged over both channels (left column). The first and second rows correspond to two representative choices of the smallest near-zero singular values that yield qualitatively distinct results.}
    \label{fig:FPGA}
\end{figure*}

\section{Experimental Setup}
We consider the quasi-2D electromagnetic cavity shown in Fig.~\ref{fig:setup}(a) that contains $15$ randomly placed metallic cylinders. This disordered system is coupled via matched coax-to-waveguide transitions to two channels at the left and right sides. Using a vector network analyzer, we can conveniently measure the $2\times2$ scattering matrix between 8 and 9~GHz (frequency step:~$\delta\nu = 1.5625$~MHz). In order to inject spatio-temporally sculpted signals in situ, we generate baseband signals between 0 and 1~GHz on an FPGA board at a sampling rate of 4 GS/s and multiply them with a local oscillator at $f=8$~GHz (see SM for additional details). We coherently record the spatio-temporal output signals with an ultra-wideband multi-channel oscilloscope using a sampling rate of 40~GS/s. 

For the particular choice of $\Delta t_x = \Delta t_y = [0~t_c]$ with $t_c= 1/(2\delta\nu)$
, we display in Fig.~\ref{fig:setup}(d) the singular value (SV) spectrum of $H$ for our disordered system shown in Fig.~\ref{fig:setup}(a). The largest SV $\lambda_1 = 0.84$ remains below unity due to the inevitable presence of absorption in our experimental system. The presence of absorption is also evidenced by the unitarity deficit of $S$ that is seen in Fig.~\ref{fig:setup}(b). This unitarity deficit is frequency dependent and it is notable in Fig.~\ref{fig:setup}(b) that for some frequencies the smaller one of the two eigenvalues of $S^\dagger S$ even gets close to zero, indicating that a time-harmonic state yielding a rather small reflected intensity exists. For the transient regime, we observe in Fig.~\ref{fig:setup}(d) that multiple of the smallest SVs of ${H}$ are extremely close to zero, suggesting that in the transient regime (i.e., within the interval $\Delta t_y$), the reflected signal intensity can be suppressed very efficiently if the corresponding spatio-temporal states are injected.

\section{Reflectionless Transient Excitation.}
We begin by considering  reflectionless transient excitation (RTE) of a scattering system. As mentioned above and seen in Fig.~\ref{fig:setup}(d), multiple $\lambda_n$ are extremely close to zero for $\Delta t_x = \Delta t_y = [0~t_c]$. We observe that two qualitatively different types of such RTEs exist in our system, and we display one experimentally measured example of each in the first and second rows of Fig.~\ref{fig:FPGA}.

The first type, in the first row of Fig.~\ref{fig:FPGA}, is an instance of a concept previously introduced as ``coherent virtual absorption''(CVA)~\cite{Baranov2017optica}. CVA is a promising concept for energy storage \cite{Trainiti2019}, sensing and transient application of optical forces \cite{Lepeshov2020} where an isolated complex-valued zero of the scattering response is engaged in the transient regime to suppress reflection within the excitation interval. Any scattering response can be decomposed into its constituent scattering singularities, namely poles $\tilde{\nu}_n$ and zeros $z_n$~\cite{grigoriev2013optimization,grigoriev2013singular,Krasnok2019}. If a zero is real-valued, reflectionless time-harmonic excitation of the scattering structure is possible~\cite{chong2010coherent,Pichler2019,Sweeney2020}. Within a limited frequency interval, a disordered system like ours usually does not have a real-valued zero unless the scattering system is purposefully perturbed to impose one~\cite{imani2020perfect,frazier2020wavefront,delHougne2020CPA,del2021coherent,sol2021meta,sol2022reflectionless}. However, given only complex-valued zeros, reflectionless excitation can still be achieved, albeit only in the transient regime for a limited temporal interval, by shaping the incident signal in space and time. When the incident signal is interrupted, the energy accumulated within the medium leaks out through the channels.

Previous experimental observations of RTE~\cite{Trainiti2019,delage2022experimental,delage2023reflectionless} considered systems with isolated resonances (similar to the first row in Fig.~\ref{fig:FPGA}). In the simple picture of isolated resonances experiencing uniform absorption within the medium, poles $\tilde{\nu}_n$ and zeros $z_n$ are found at the same real frequency $\nu_n$ but acquire different imaginary parts, $\tilde{\nu}_n = \nu_n-i(\gamma_n+\gamma_a)$ and $z_n=\nu_n+i(\gamma_n-\gamma_a)$ ($\gamma_n$ is the resonance width and $\gamma_a$ is the absorption strength). The incident signal corresponding to a zero $z_n$ is a monochromatic signal at frequency $\mathrm{Re}[z_n]$ modulated with an exponentially increasing envelope at the rate $\mathrm{Im}[z_n]$~\cite{Baranov2017optica}. Identifying poles and zeros however requires cumbersome protocols that becomes highly challenging in the regime of overlapping resonances~\cite{Trainiti2019,delage2022experimental,delage2023reflectionless,chen2022use,Mandelshtam1997,Kuhl2008,Davy2018}. 
In contrast, our technique solely requires the SVD of the transport matrix $H$ and directly yields the non-intuitive spatio-temporal input signals. For overlapping resonances, these signals involve multiple zeros as seen in the second row of Fig.~\ref{fig:FPGA}.



How many VPA states exist for a given transport matrix $H$? Theoretically, we expect this number to correspond to the number of complex zeros in the upper half of the complex frequency plane. We show in the SM that a VPA state indeed disappears as the corresponding zero crosses the real axis and acquires a negative imaginary part. The total number of poles and zeros can be estimated using Weyl's law~\cite{Weyl1911,Arendt,Pierrat2014}. The average density of states $\rho(\omega) \sim A\omega / (2 \pi c_0^2)$, where $A$ is the area of the scattering region, gives a theoretical estimate of 25 modes between 8 and 9~GHz. This number can also be roughly confirmed by counting the peaks in Fig.~\ref{fig:setup}(b). However, counting the number of zeros with positive imaginary part is more challenging. To do so, we study in Fig.~\ref{fig:setup}(c) the number of positive peaks of the real part of the Wigner-Smith time delay $\tau_W(\nu) = [\mathrm{Tr}(Q(\nu))]$. Here, $Q(\nu)=-i \left[ S(\nu) \right]^{-1}  [\partial S(\nu)/ \partial \nu]$ is the Wigner-Smith operator~\cite{Wigner1955,Smith1960}. The real part of the eigenvalues of $Q$ is assumed to be proportional to a dwell time~\cite{fan2005principal,Rotter2011,Davy2015,Boehm2018,Durand2019,delHougne2021,huang2022wave} which is diverging as a zero crosses the real axis  \cite{Asano2016,delHougne2020CPA,ChenPRL2021,ChenPRE2021}.
A positive (negative) peak of $\mathrm{Re}[\tau_W]$ is a clear signature of a zero with positive (negative) imaginary part. The number of positive peaks (16) in $\mathrm{Re}[\tau_W(\nu)]$ is in good agreement with the number of $\lambda_n$ below $\lambda = 10^{-5}$ (15). This number increases with the number $N$ of ports connected to the cavity as the channel decay rate scales as $\langle \gamma_n \rangle = N \gamma_0 + \gamma_a$, where $\gamma_0$ is the single-channel decay rate. In our system, if we increase $N$ from 2 to 6, $\gamma_n \gg \gamma_a$ so that all the zeros have positive imaginary parts and the number of VPA states is equal to 25 (see SM).  

\section{Optimal Energy Delivery}

\begin{figure}
    \centering
    \includegraphics[width=8.5cm]{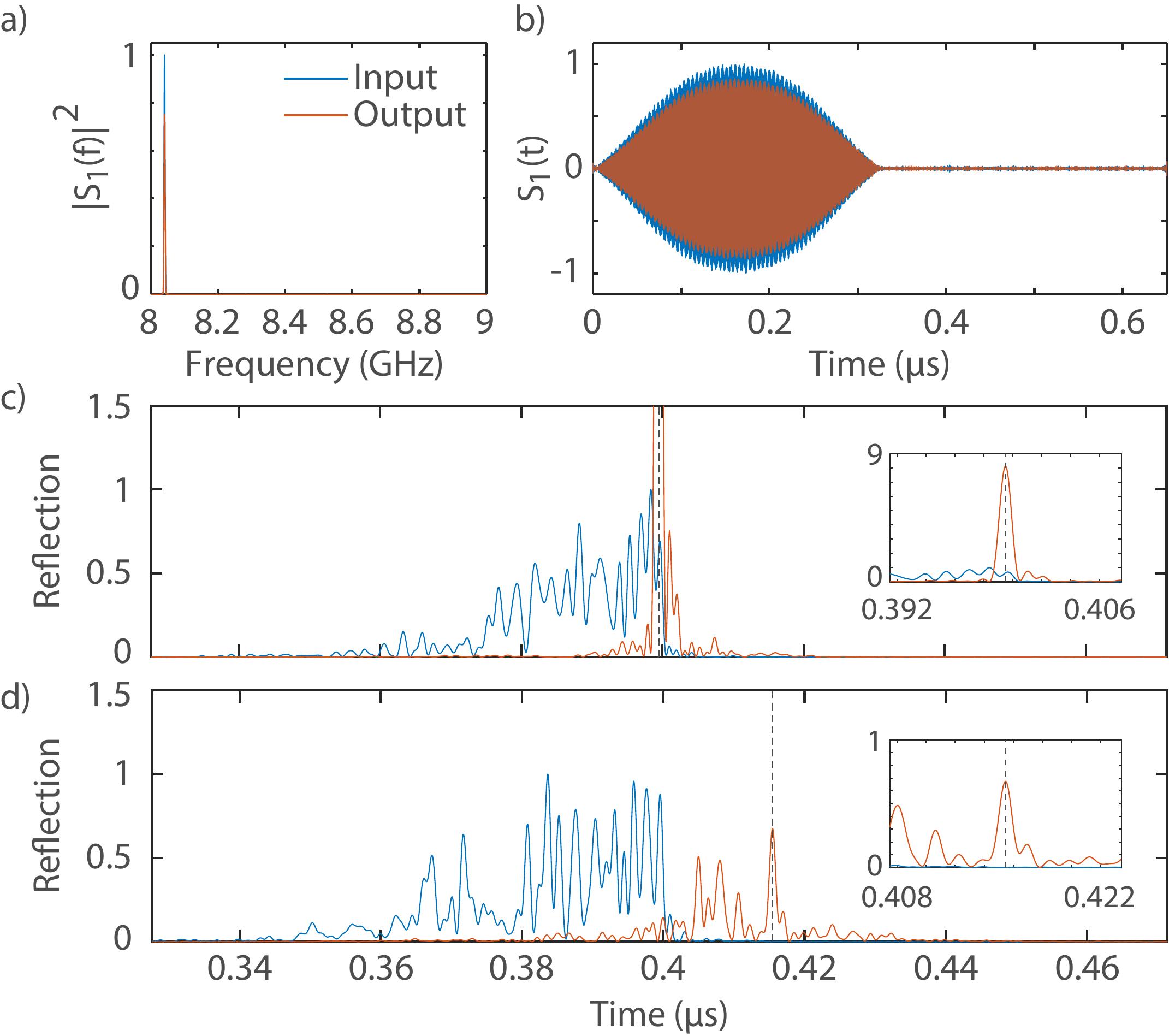}
    \caption{(a,b) Optimal deposition of energy corresponding to the largest singular value of 0.84 for $\Delta t_x = \Delta t_y = [0 ~ 0.5 t_{\mathrm{max}}]$ with $t_{\mathrm{max}} = 0.64$~$\mu$s. (c,d) Optimal energy deposition at two chosen output times: (c) $t = 0.5 t_{\mathrm{max}}$, (d) $t = 0.52 t_{\mathrm{max}}$ with $t_{\mathrm{max}} = 0.80$~$\mu$s. The last two rows show the spatially averaged time-varying input (blue) and output (red) intensities upon injecting states for maximal reflection corresponding to the first singular value of $H$. The insets are a zoom around the focal time.
    }
    \label{fig:arbitrary_time}
\end{figure}

\begin{figure}
    \centering
    \includegraphics[width=8.5cm]{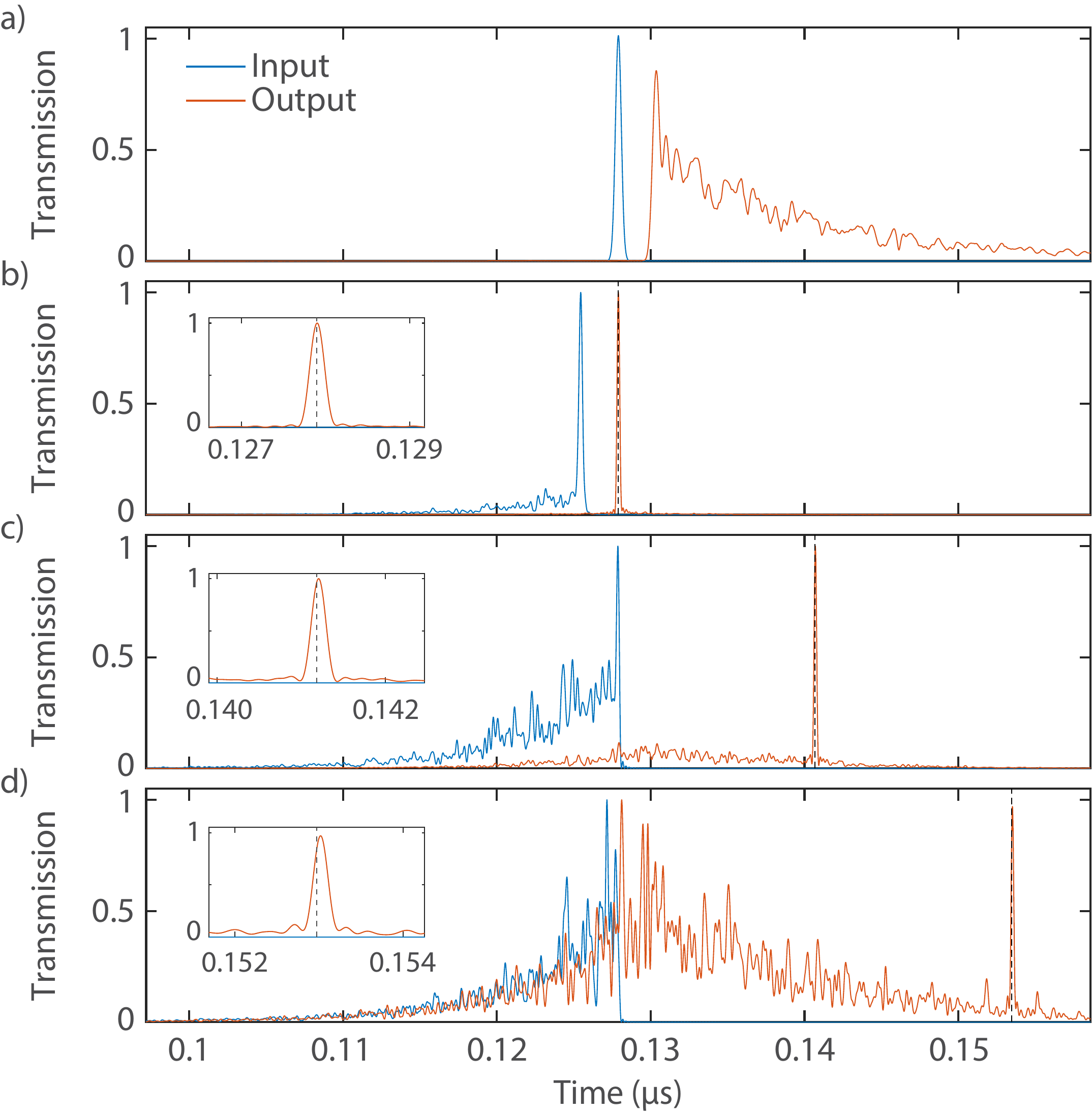}
    \caption{(a) Temporal transmitted intensity through the cavity for an incident pulse at $t = 0.5 t_{\mathrm{max}} = 0.128~\mu$s transmitted on a single channel. (b-d) Optimal energy deposition at $t_c = 0.5 t_{\mathrm{max}}$ (b), $t_c = 0.55 t_{\mathrm{max}}$ (c) and $t_c = 0.6 t_{\mathrm{max}}$ (d) with $t_{\mathrm{max}} = 0.26$~$\mu$s.  }
    \label{fig:Transmission Focusing}
\end{figure}

We now turn our attention to the optimal energy deposition, which consists in delivering the maximum possible amount of energy within the output time interval $\Delta t_y$ given spatio-temporal coherent control in the input time interval $\Delta t_x$. In Fig.~\ref{fig:arbitrary_time}(a,b), we experimentally observe the input and output signals upon injecting the state corresponding to the largest singular value of ${H}$ for the intervals considered in Fig.~\ref{fig:setup}(d). During the transient excitation, a very large fraction ($\lambda_1 = 0.84$) of the injected energy exits the system within the same temporal interval ($\Delta t_x = \Delta t_y$). The remainder of the injected energy is absorbed in the system. Maximal-reflection states corresponding to other temporal intervals are presented in the SM.

By adjusting the definition of the output time interval $\Delta t_y$, the same approach can also yield optimal energy delivery at arbitrary times after the transient excitation interval. Two examples in the limit of a short output interval $\Delta t_y = \delta(t-t_c)$ are shown in Fig.~\ref{fig:arbitrary_time}(c,d). In all cases, a sharp increase of total reflection at the selected time is observed. Of course, the duration of the focused output signals is not arbitrarily short but is instead determined by the bandwidth of the spatio-temporal input signal. 

The number of transmitting channels is here limited ($N=2$) by our electronic hardware. To further demonstrate the interest of our approach, we consider a second 2D cavity of length $L = 0.5$~m and width $W = 0.25$~m with two arrays of $N = 8$ coax-to-waveguide transitions attached on the left and right interfaces (see Ref.\cite{Davy2021mean} for a description of the experimental setup). We measure the transmission matrix (TM) in the diffusive regime between  between 10 and 15~GHz with a frequency step of 3.9~MHz. For a short incident pulse delivered at $t = 0.5 t_{\mathrm{max}}$ through a single channel, the temporal profile of the output intensity decays exponentially with a decay time equal to $\tau = 0.0186 t_{\mathrm{max}}$ ($t_{\mathrm{max}} = 0.26$~$\mu$s), see Fig.~\ref{fig:Transmission Focusing}(a). Using the linearity of the wave equation, we reconstruct the transmitted intensity averaged over outgoing channels for an optimal energy delivery at time $\Delta t_y = \delta(t-t_c)$ with $t_c = 0.5 t_{\mathrm{max}}$, $t_c = 0.55 t_{\mathrm{max}}$ and $t_c = 0.6 t_{\mathrm{max}}$ (see Fig.~\ref{fig:Transmission Focusing}(b-d)). In each case the optimal incoming signal is assumed to be injected within the interval $\Delta t_y = [0-0.5t_{\mathrm{max}}]$. A clear enhancement of the transmitted  signals at selected temporal interval is now observed even at late times.

Our spatio-temporal matrix formulation takes full advantage of both spectral and spatial DoFs to achieve optimal total transmission at selected times.  We show in Fig.~\ref{fig:Transmission Focusing} that the transmitted intensity at the focal time scales linearly with the number of incoming antennas $N$ and the bandwidth of incident signals $\Delta \nu$. This is in agreement with theoretical predictions for time-reversal experiments in diffusive or chaotic systems, which demonstrates that the focused intensity increases linearly with the number of DoFs $N \times N_f$. Here $N_f$ is the ratio between the bandwidth $\Delta \nu$ and the spectral correlation length $\delta \nu \sim 1/\tau$. For $N=8$ and a bandwidth of 5~GHz, we estimate that the total DoFs is equal to 190, which is of the same order of magnitude as the number of spatially controllable optical modes in Ref.~\cite{devaud2022temporal}. 


\begin{figure}
    \centering
    \includegraphics[width=8.5cm]{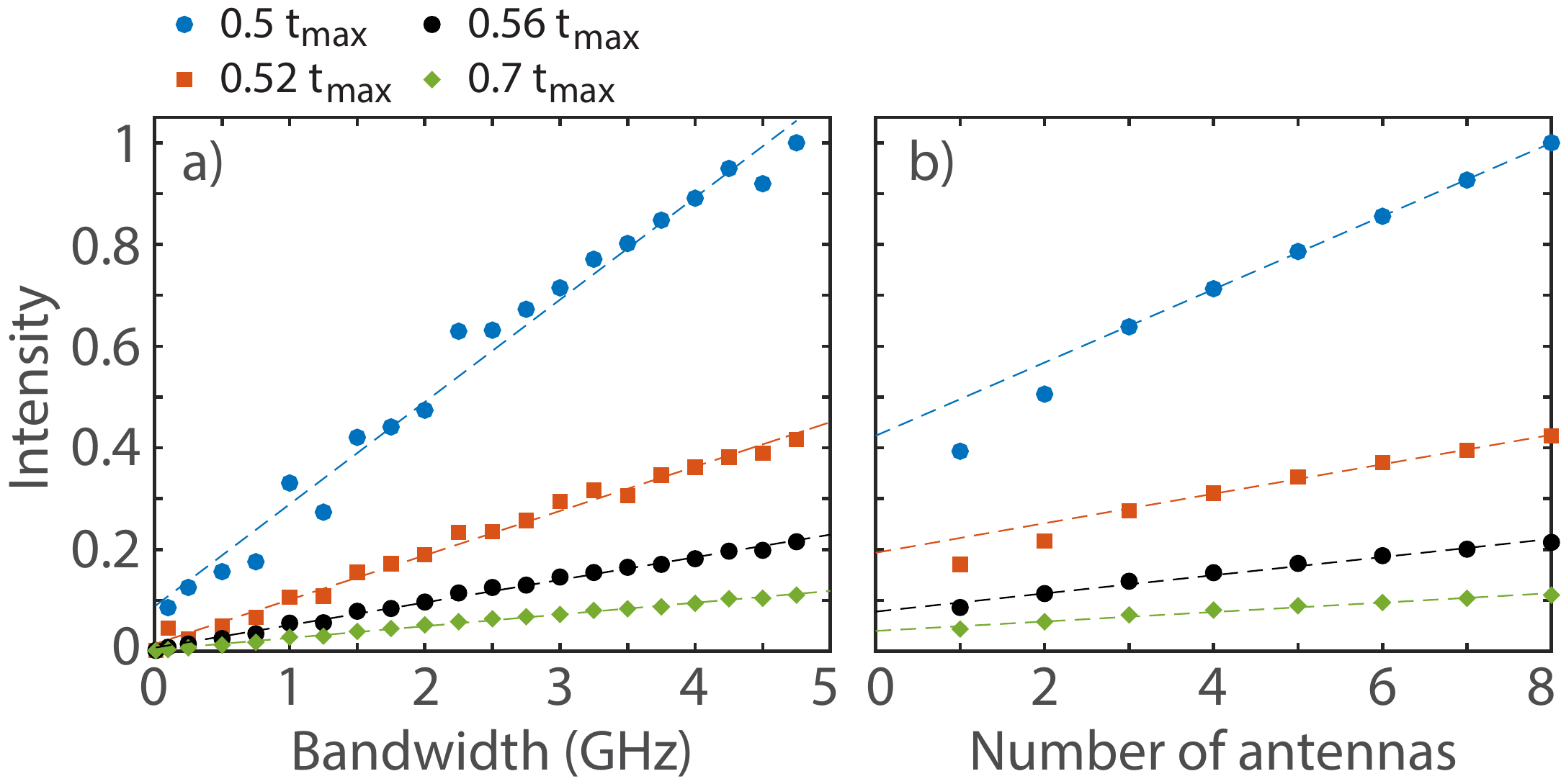}
    \caption{Enhancement of transmitted intensity with respect to the bandwidth (a) and number of antennas (b) used for optimal transmitted energy. The markers represent experimental data for different selected times $t_c$ between $t_c = 0.5 t_{\mathrm{max}}$ and $t_c = 0.7 t_{\mathrm{max}}$ with $t_{\mathrm{max}} = 0.26$~$\mu$s. The dashed lines are linear fits of the data. The intensity is normalized by its result found for $N=8$, a bandwidth of 5~GHz and $t_c = 0.5 t_{\mathrm{max}}$}
\end{figure}

\section{Scattering-Invariant Time-Varying States.}

\begin{figure}
    \centering
    \includegraphics[width=8.5cm]{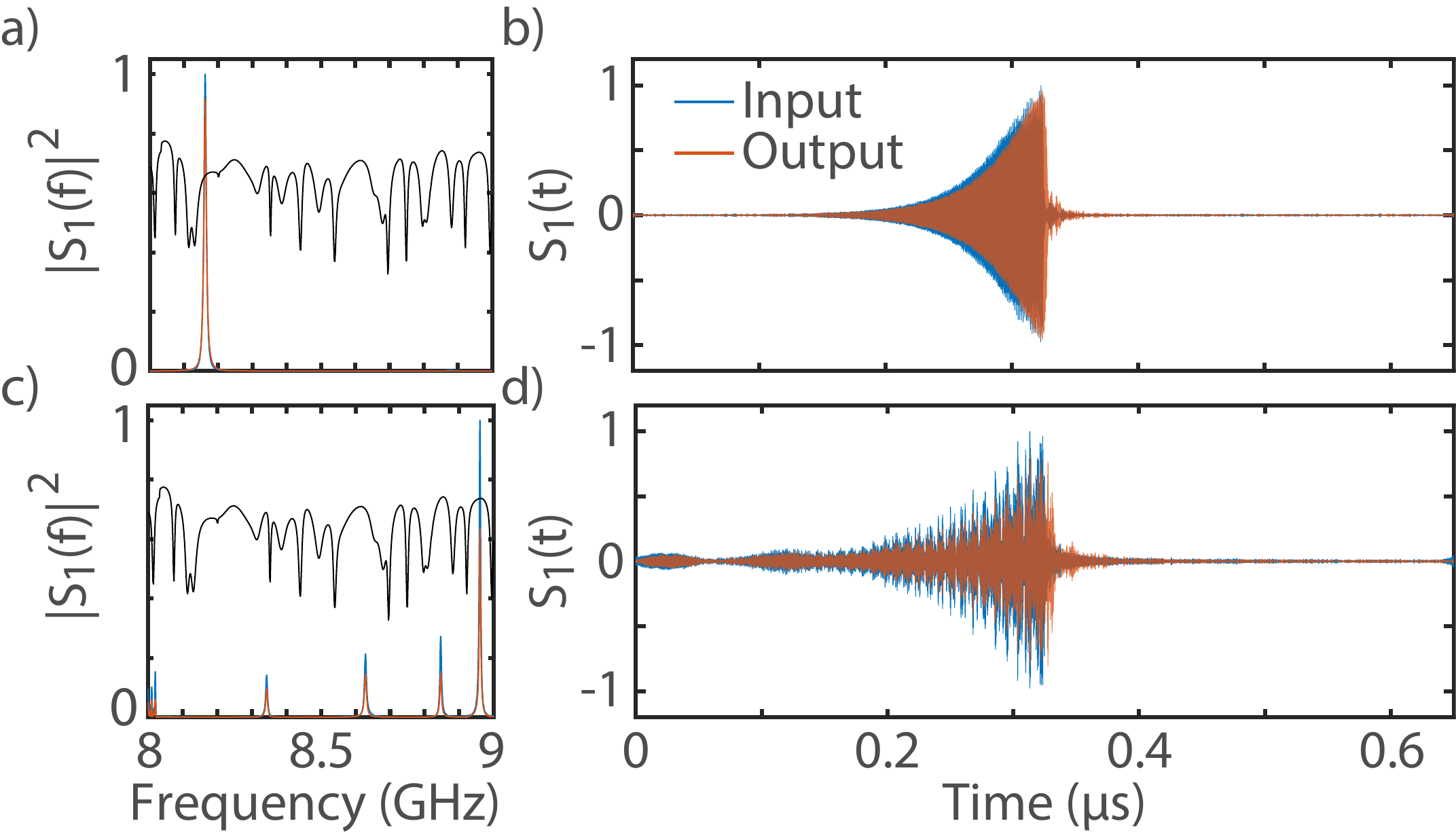}
    \caption{Spectral (a,c) and temporal (b,d) representation of two scattering-invariant time-varying states for a temporal interval of $[0 \ 3.18]$~$\mu$s (i.e. $[0 \ t_c]$ with $t_c=2/\delta\nu$). The input (blue) and output (red) signals are shown for the first channel (and look similar for the second channel). The eigenvalues are $|\alpha|= 0.71$ in (b) and $|\alpha| = 0.58$ in (d). The black curve in (a,c) is the reflection $R(\nu)$ averaged over the input ports.}
    \label{fig:invariant}
\end{figure}

Finally, we explore the possibility of exciting states that have the same spatial and temporal input and output patterns, generalizing the recent concept of time-harmonic (monochromatic) scattering-invariant states that have the same spatial input and output patterns~\cite{Pai2021}. To that end, we solve the eigenvalue problem 
\begin{equation}
   {H} x_n = \alpha_n x_n 
\end{equation}
and extract the input signals giving optimal similarity to the output signals within an arbitrary interval. States with smallest $|\alpha_n|$ obviously correspond to low-reflection transient excitations. Of more interest are states with maximal $|\alpha_n|$. Our approach guarantees that the similarity between input and output states is also maximal. We have injected in the smaller cavity with $N=2$ two such scattering-invariant time-varying states with large eigenvalue $|\alpha_n|$, $|\alpha| = 0.71$ and $|\alpha| = 0.58$ see Fig.~\ref{fig:invariant}. The input and output temporal intervals are $\Delta t_x = \Delta t_y = [0~ 0.5t_{\mathrm{max}}]$. Their spectra are peaked on a single or a few frequencies with a bandwidth inversely proportional to the temporal interval. In the time domain, both signals are increasing exponentially. While RTE states are associated with seemingly large delay times, the scattering-invariant time-varying states require quasi-non-dispersive transport and correspond to short delay times. They are therefore peaked within frequency ranges in which the reflection coefficient presents a flattened shape [see Fig.~\ref{fig:invariant}(a,c)].  

\begin{figure}
    \centering
    \includegraphics[width=8.5cm]{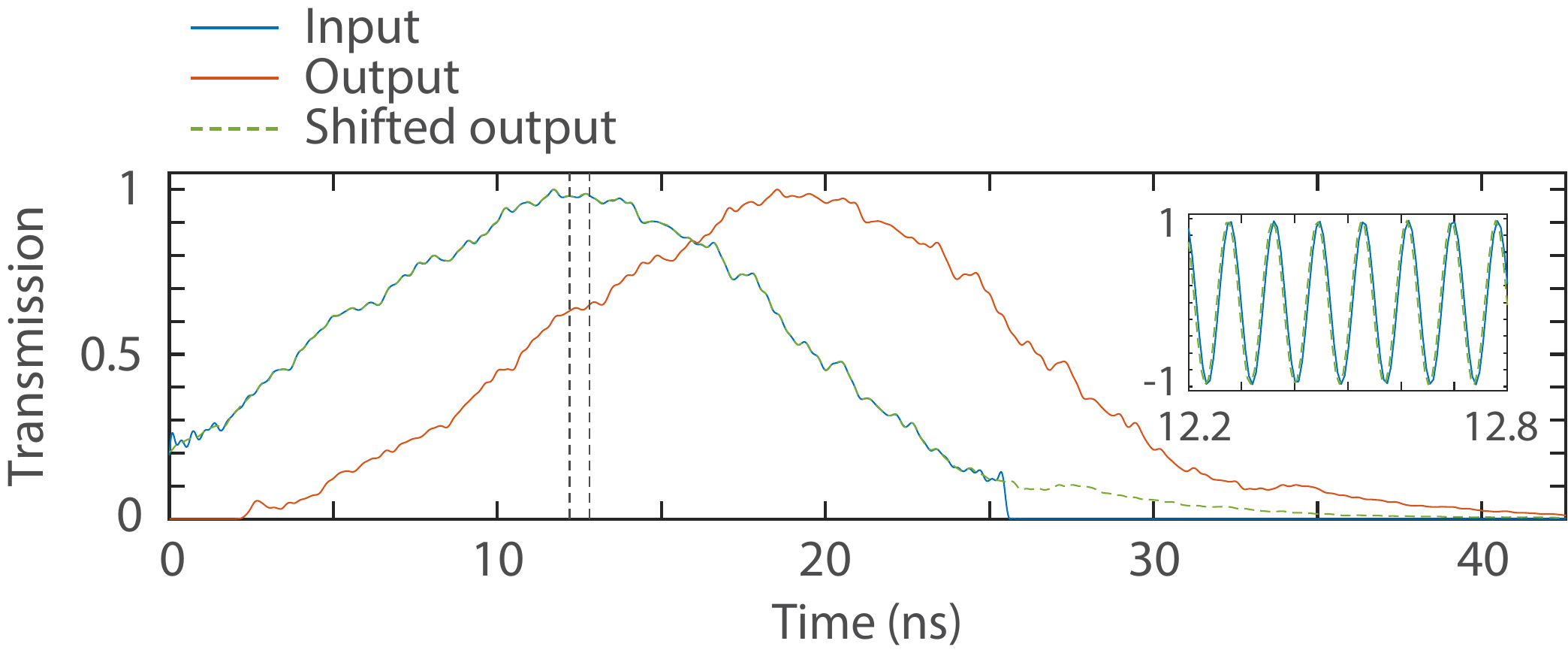}
    \caption{Temporal intensity profile of an invariant time-varying state in transmission through the large cavity for an incident interval $[0 ~ 25]$ns and an output interval $[8 ~ 0.33 ]$ns. The blue line is the incident signal, the red line is the outgoing signal and the green dashed line is the output signal shifted by $-\Delta \tau = -8$ns. The inset is a zoom of the field on a small interval represented by the dashed black lines in the figure.}
    \label{fig:Invariant_Transmission}
\end{figure}

We then consider scattering-invariant time-varying states in transmission through the larger cavity with $N=8$ antennas on each side. The length of the input interval is fixed to $0.3 t_{\mathrm{max}}$ and the output temporal interval is now shifted by $d\tau$, $\Delta t_y = \Delta t_x + \Delta \tau$. We find a maximal similarity coefficient $|\alpha| = 0.77$ for $\Delta \tau = 0.08 t_{\mathrm{max}}$. The corresponding length $c_0 \Delta\tau = 2.04$~m is close to the mean path length of the cavity~\cite{Davy2021mean}. The average temporal envelopes of input and output signals are shown in Fig.~\ref{fig:Invariant_Transmission}. The output signal shifted by $-\Delta \tau$ is very similar to the injected signal. Since input and output speckle patterns are correlated over a temporal interval, these states could be useful for broadband imaging through disordered systems.

\section{Conclusion}
We have presented a matrix formalism enabling optimal spatio-temporal coherent control of waves in arbitrarily complex scattering systems by taking full advantage of both spatial and temporal DoFs. Our work presents a unifying perspective on many contemporary wave-control techniques (time reversal, wavefront shaping, virtual perfect absorption, scattering-invariant modes) and even generalizes them. Beyond providing fundamental insights, our technique may readily find applications in the microwave and acoustics regimes where the necessary hardware for coherent spatio-temporal wavefront generation is available.

\subsection*{Acknowledgements}
\noindent We acknowledge Clara Moy for assistance with the work presented in Sec.~IV. This publication was supported by the European Union through the European Regional Development Fund (ERDF), by the French region of Brittany and Rennes M{\'e}tropole through the CPER Project SOPHIE/STIC \& Ondes. M. D. acknowledges the Institut Universitaire de France. C. F. acknowledges funding from the French ``Minist{\`e}re de la D{\'e}fense, Direction G{\'e}n{\'e}rale de l'Armement''.

\bibliographystyle{apsrev4-1}

\providecommand{\noopsort}[1]{}\providecommand{\singleletter}[1]{#1}%
%

\clearpage

\onecolumngrid

\begin{center}
\textbf{\large Supplemental Material for \\``Matrix approach for optimal spatio-temporal coherent control of wave scattering''} \\ [.4cm]
  Clément Ferise,$^{1}$ Philipp del Hougne,$^{1}$ and Matthieu Davy$^1$ \\[.1cm]
  {\itshape ${}^1$Univ Rennes, CNRS, IETR - UMR 6164, F-35000 Rennes, France}\\
  (Dated: \today) \\ [1cm]
\end{center}

\twocolumngrid

\setcounter{equation}{0}
\setcounter{figure}{0}
\setcounter{table}{0}
\setcounter{page}{1}
\setcounter{section}{0}
\makeatletter
\renewcommand{\theequation}{S\arabic{equation}}
\renewcommand{\thefigure}{S\arabic{figure}}
\renewcommand{\bibnumfmt}[1]{[S#1]}
\renewcommand{\citenumfont}[1]{S#1}

\section{Theory}

We provide here details on the multichannel temporal operator given in the main text. We consider $N$ time signals $x_n(t_m)$ sampled on $N_{tx}$ points between $0$ and $t_{c}$. The Fourier transform on $N_\nu$ frequencies
\begin{equation}
    S(\nu_k) = \frac{1}{\sqrt{N_{tx}}} \displaystyle\sum_{m=0}^{N_{tx}-1}s(t_m)e^{-2i\pi \nu_k t_m}
\end{equation}
can be expressed in terms of the discrete Fourier transform (DFT) operator, $x_n(\nu) = D_x x_n(t) $ with elements

\begin{equation}
    [D_x]_{km} = \frac{1}{\sqrt{N_{tx}}}e^{-i 2 \pi \nu_k t_m }.
\end{equation}
The dimensions of $D_x$ is $N_\nu \times N_{tx}$. When we set $\delta t = 1/(N_\nu d\nu)$ and $t_c = t_{\mathrm{max}}$ with $t_{\mathrm{max}} = 1/\delta\nu$, the operator $D_x$ is unitary, meaning that $D_xD_x^\dagger = \id_{N_\nu}$. In the more general case of control over smaller time invervals $t_c < t_{\mathrm{max}}$ ($N_{tx} < N_\nu$), the DFT operator is non-square. It however still verifies ${D}_x^\dagger {D}_x = \id$ but ${D}_x {D}_x^\dagger \neq \id$.

We then seek for a single multichannel linear relation for the $N$ input and output channels. We first concatenate the spectral and temporal matrices into the single vectors $x_t$ and $x_\nu$ of length $NN_{tx}$ and $NN_{\nu}$, given below for $N=2$:
\begin{equation}
    x_t = \begin{pmatrix}
    x_{1}(t_1)\\
    x_{2}(t_1)\\
    x_{1}(t_2)\\
    x_{2}(t_2)\\
    \vdots\\
    x_{1}(t_{N_{tx}})\\
    x_{2}(t_{N_{tx}})
    \end{pmatrix} ~ \textnormal{and} ~       x_f = \begin{pmatrix}
    x_{1}(f_1)\\
    x_{2}(f_1)\\
    x_{1}(f_2)\\
    x_{2}(f_2)\\
    \vdots\\
    x_{1}(f_{N_{\nu}})\\
    x_{2}(f_{N_{\nu}})
    \end{pmatrix}
\end{equation}

The elements of $x_t$ are therefore $x_{t_j} = x_n(t_m)$ with $j = (m-1)N + n$. Using this formulation, we can find a linear relation between $x_t$ and $x_\nu$. We introduce the sparse multichannel DFT operator $\tilde{D}_x$ of dimensions $NN_\nu \times NN_{tx}$ expressed as (still for $N=2$):

\begin{equation}
\label{D}
    \tilde{D}_x = \begin{pmatrix} (D_x)_{00} & 0 & \cdots & (D_x)_{0N_{tx}} & 0 \\ 0 & (D_x)_{00} & \cdots & 0 & (D_x)_{0N_{tx}} \\ (D_x)_{10} & 0 &  \cdots & (D_x)_{1N_{tx}} & 0 \\ 0 & (D_x)_{10} &  \cdots & 0 & (D_x)_{1N_{tx}} \\ \vdots & \vdots  & \ddots & \vdots & \vdots \\ (D_x)_{N_\nu0} & 0 & \cdots & (D_x)_{N_\nu N_{tx}} & 0 \\ 0 & (D_x)_{N_\nu 0} &  \cdots & 0 & (D_x)_{N_\nu N_{tx}} \end{pmatrix} 
\end{equation}
to get
\begin{equation}
    x_\nu = \tilde{D}_x x_t
\end{equation}
One can easily verify that $\tilde{D}_x$ has the same properties as $D_x$: it is unitary for $t_c = t_{\mathrm{max}} $ and that $\tilde{D}_x^\dagger \tilde{D}_x = \id$ for any temporal interval. However $\tilde{D}_x \tilde{D}_x^\dagger \neq \id$ for $t_c < t_{\mathrm{max}} $.

We apply the same procedure to the $M$ outgoing signals $y_n(t)$ measured within the time window $\Delta t_y$. Finally, we define the block diagonal matrix $S_D $ with 
\begin{equation}
    S_D =\begin{pmatrix}
    S(\nu_1) & 0 & 0 & \cdots & 0\\
    0 & S(\nu_2) & 0 & \cdots & 0\\
    \vdots & 0 & \ddots & 0 & \vdots\\
    \vdots & \vdots & 0 & \ddots & \vdots\\
    0 & 0 & \cdots & 0 & S(\nu_{N_\nu})
    \end{pmatrix}
\end{equation}
We finally get $\tilde{D}_y y_t = S_D \tilde{D}_x x_t$ and
\begin{equation}
    y_t = \tilde{D}_y^\dagger S_D \tilde{D}_x x_t.
\label{eq:linear}
\end{equation}
which is Eq.~(1) of the main text.

\section{Absorption and virtual perfect absorption}

\begin{figure}
    \centering
    \includegraphics[width=8.5cm]{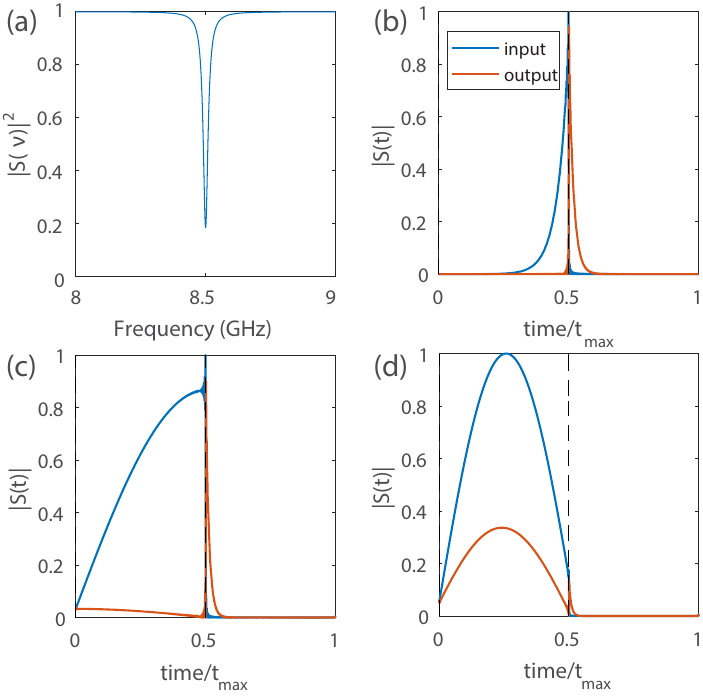}
    \caption{(a) Reflection $|S(\nu)|^2$ for a single resonance at $\nu_n = 8.5$~GHz. The scattering matrix is described by the Breit-Wigner formula given in Eq.~(\ref{eq:Breit}). The channel decay rate is $\gamma_n = 8$~MHz and $\gamma_a = \gamma_n /2$. (b-d) Input and output signals for the minimal subspace of $H$ corresponding to an excitation between $0$ and $t_c = t_{\mathrm{max}}/2$ ($t_{\mathrm{max}}/2 = 1/\delta\nu$) for $\gamma_a = \gamma_n/2$ (b), $\gamma_a = \gamma_n$ (c) and $\gamma_a = 2\gamma_n$ (c) }
    \label{fig:zero}
\end{figure}

\begin{figure}
    \centering
    \includegraphics[width=8.5cm]{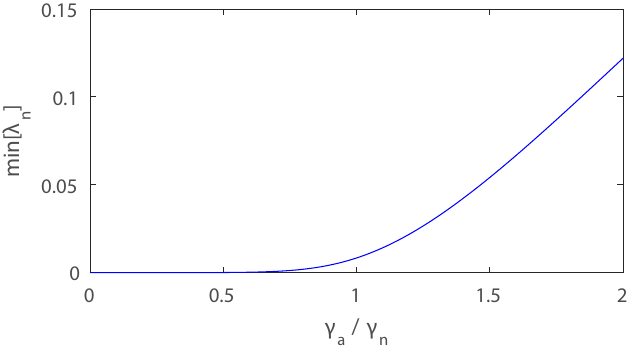}
    \caption{Variation of the last eigenvalue of $H\dagger H$ as a function of the absorption rate $\gamma_a / \gamma_n$. VPA states with suppressed reflection are obtained only as $\gamma_a \ll \gamma_n$.}
    \label{fig:zero_gamma_a}
\end{figure}

In this section, we demonstrate numerically that the VPA states disappear when a zero of the scattering matrix $S$ crosses the real axis. 
We consider in the following a single channel $N=1$ system but an extension to multichannel systems is straightforward. For a single resonance, the scattering matrix is expressed by a Breit-Wigner formula,
\begin{equation}
    S(\nu) = 1-\frac{i\gamma_n}{\nu - \nu_n +i(\gamma_n+\gamma_a)} 
\label{eq:Breit}
\end{equation}
Here $\gamma_n$ represents the coupling of the channel to the system and $\gamma_a$ is the absorption linewidth. The reflection coefficient $|S(\nu)|^2$ is shown in Fig.~\ref{fig:zero}(a) for $\gamma_a = \gamma_n/2$. The resonance is clearly visible at $\nu_n = 8.5$~GHz.

The pole of $S$ is $\tilde{\nu}_n = \nu_n -i(\gamma_n+\gamma_a)$ and the zero of $S$ is $z_n = \nu_n +i(\gamma_n-\gamma_a)$. When $\gamma_n > \gamma_a$, the imaginary part of $z_n$ is positive and the zero is located in the upper complex plane. This zero can then be accessed by sending a monchromatic signal at frequency $\nu_n$ modulated by an exponentially growing envelope at rate $\gamma_n - \gamma_a$. This input temporal signal $x(t)$ for an excitation between $0$ and $t_c = t_{\mathrm{max}}/2$ is directly extracted from the last subspace of the corresponding $H = \tilde{D}_y^\dagger S_D \tilde{D}_x $ matrix. The outgoing energy in such state of virtual perfect absorption is suppressed between $0$ and $t_c$. Once the excitation is turned off, the energy stored within the cavity that has not been absorbed is released at rate $\gamma_n$. The envelopes of input and output signals are illustrated in Fig.\ref{fig:zero}(b) for $\gamma_a = \gamma_n/2$. 

However, as $\gamma_a$ increases, zero reflection between $0$ and $t_c$ becomes impossible for an excitation only within this range. For instance, for $\gamma_a = \gamma_n$, the zero of $S$ is real, $z_n = \nu_n$, and the excitation $x(t)$ must be fully monochromatic to be perfectly absorbed with no reflection. This is the condition of coherent perfect absorption found when the absorption decay rate balances the excitation rate~\cite{chong2010coherent}. The reflection between $0$ and $t_c$ for any non-monochromatic signal cannot be suppressed. In Fig.\ref{fig:zero}(c), we observe that the minimal reflection state of the $H$ matrix is still increasing in time but the reflection is not zero. For $\gamma_a > \gamma_n$, the state with minimal reflection presents a pronounced peak at $t = t_c/2$ [see Fig.\ref{fig:zero}(d)].

In Fig.~\ref{fig:zero_gamma_a}, we show the variation of the minimal reflection value found from the singular value decomposition of $H$. It clearly shows that the reflection can indeed be completely suppressed for $\gamma_a \ll \gamma_n$ but this property is lost as $\gamma_a$ approaches $\gamma_n$ as expected. For $\gamma_a = 2\gamma_n$, the smallest reflection is even quite strong ($\mathrm{min}[\lambda_{n}] = 0.13$).

\section{Details on the experimental setup}
\begin{figure}
    \centering
    \includegraphics[width=8.5cm]{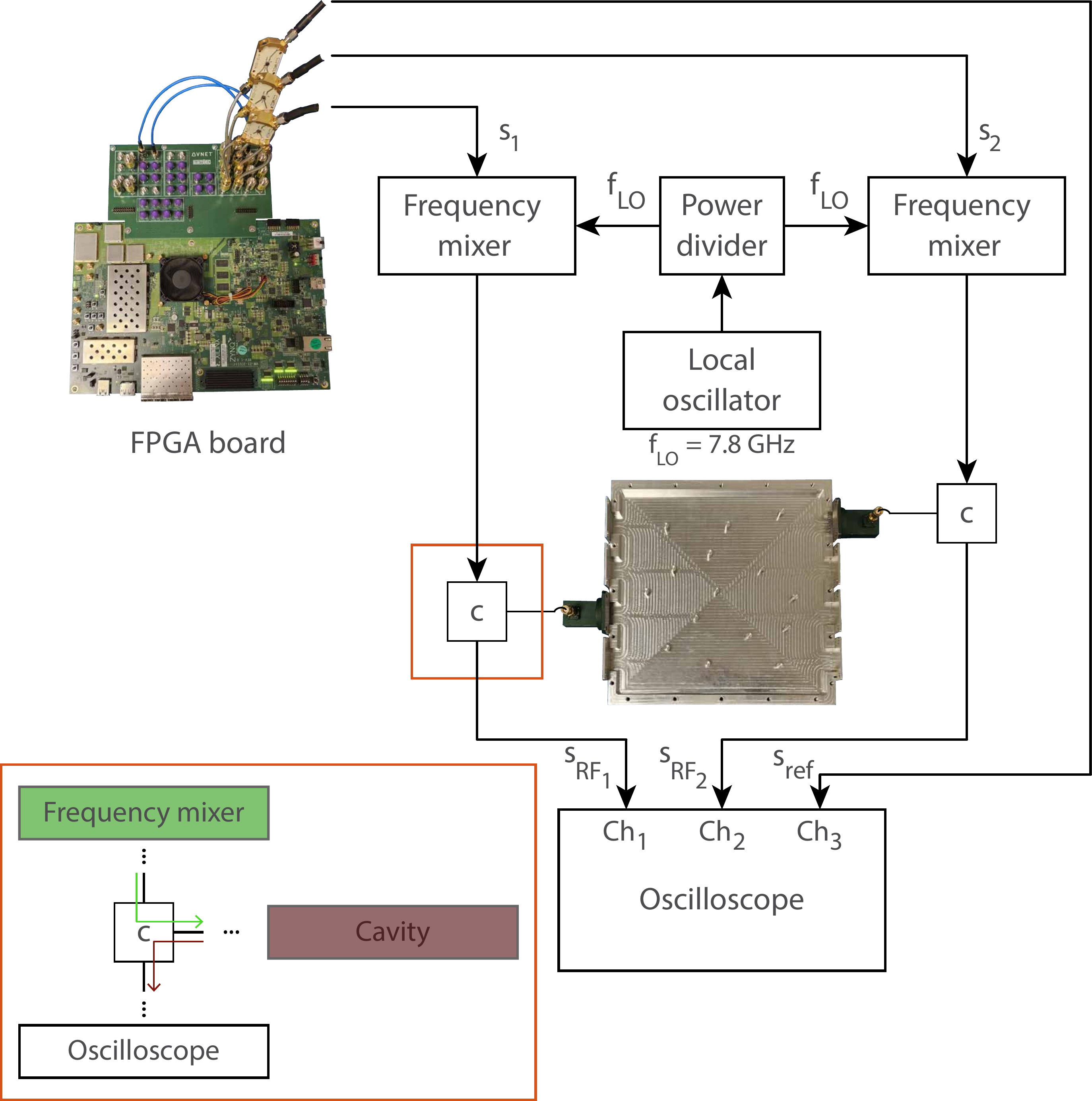}
    \caption{Detailed sketch of the experimental setup. The inset explicitly shows the role of the circulators.}
    \label{fig:SM_ExperimentalSetup}
\end{figure}

In this section we present in details our experimental setup. A sketch is shown on Fig~\ref{fig:SM_ExperimentalSetup}.
A $4 \times 4$ multi-channel square cavity (side $L=205.74$~mm and height $H=10.16$~mm) comprising only two coupled ports to the system is used. Uncoupled ports are obstructed by a metallic plate.
The measurement of the $2 \times 2$ scattering matrix between $8$ and $9$~GHz is operated with a Vector Network Analyser (VNA) on the two coupled ports of the cavity via coax-to-waveguide transitions with one circulator (Aerocomm $J80.160$) connected to it.
We use circulators on each port to inject and measure the signals with a $4$-ports VNA.

By applying the SVD-based approach and matrix formalism presented in the main text, we obtain  baseband temporal signals with frequencies between $0$ and $1$~GHz. These signals are first up-converted in the frequency range between $0.2$ and $1.2$~GHz via a Fourier transform and an inverse Fourier transform successively. The two pseudo-baseband signals ($s_1$ and $s_2$ on Fig~\ref{fig:SM_ExperimentalSetup}) are finally uploaded on a FPGA board (Xilinx \textit{ZYNQ-ZCU111}). They are analogically up-converted in the frequency range $[8~9]$~GHz using two wide band frequency mixers  (Mini-Circuits \textit{ZX05-153LH-S+}) connected to a local oscillator at $7.8$~GHz. The same local oscillator is connected to the frequency mixers via a 4 ways power divider (\textit{AMD-GROUP PD4-2-18-10}) to synchronize the signals. The RF output of the frequency mixers is connected to the two circulators matched with the two coax-to-waveguide transitions coupled to the cavity. 
To measure the signals reflected from the cavity, the third channel of each circulator is connected to an ultra-wideband oscilloscope with a sampling rate of $40$GSamples/s (\textit{SDA 816Zi-B)}. 

\section{Impact of the size of the input time interval}
\begin{figure}
    \centering
    \includegraphics[width=8.5cm]{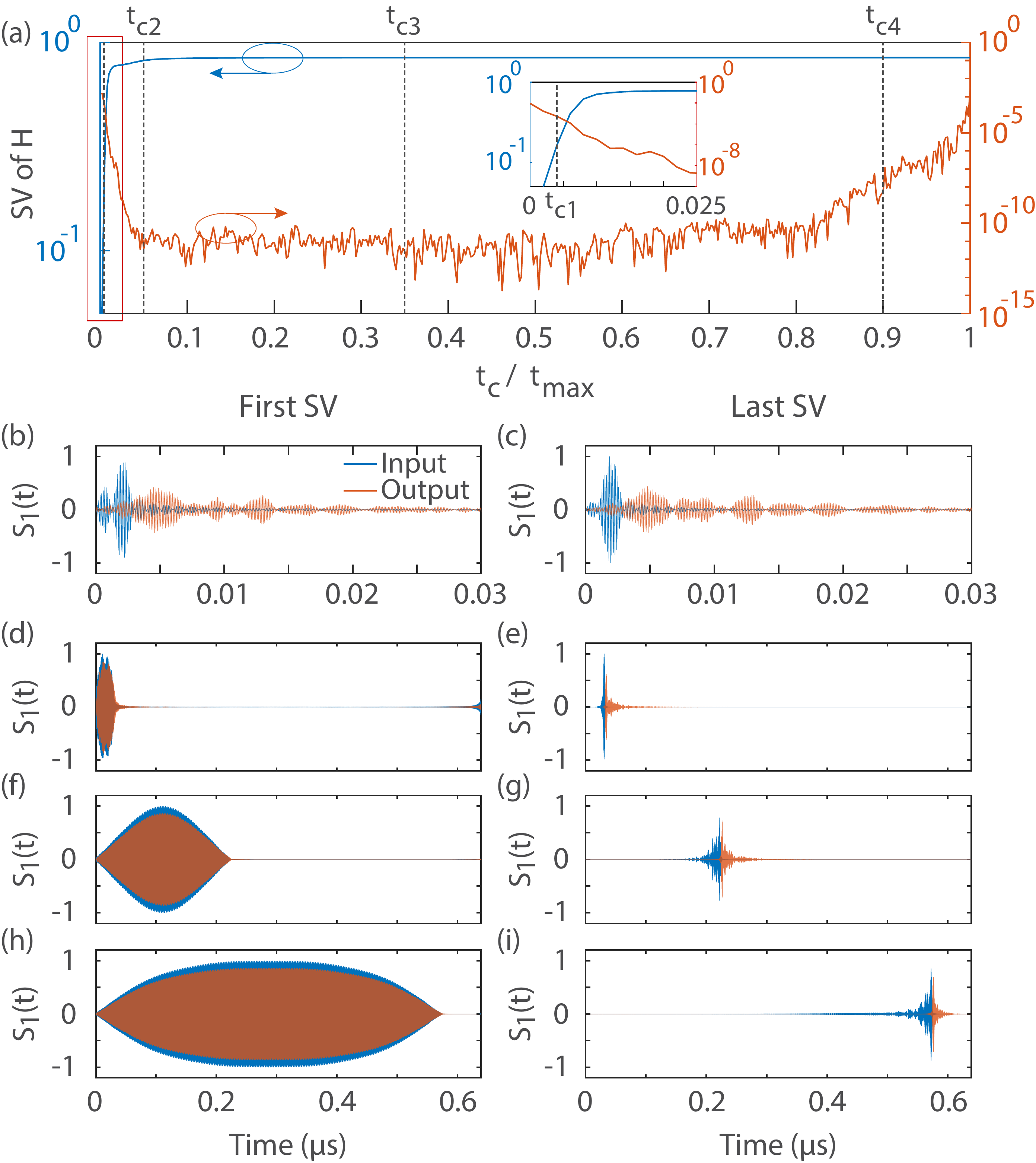}
    \caption{(a) Spectrum of the first (blue) and last (red) singular value of H as a function of the variation of $t_c$.
    (b-i) Temporal representation from numerical simulations upon coherent spatio-temporal injection of the first (b,d,f,h) and the last (c,e,g,i) right singular vectors of $H$ for four different input time intervals: (b,c) $t_{c1}=0.004t_{max}$, (d,e) $t_{c2}=0.05t_{max}$, (f,g) $t_{c3} = 0.35t_{max}$ and $t_{c4} = 0.9t_{max}$.}
    \label{fig:SM_Sweep_tc}
\end{figure}

In Fig.~2 on the main text we have presented our new approach by displaying optimal energy deposition and its opposite: reflectionless transient excitation for $\Delta t_x = \Delta t_y = [0~t_c]$ with $ t_c= t_{\mathrm{max}}/2$ and for $N=2$ channels coupled to the scattering system. In this section we numerically explore the impact of the time interval on reflection coefficients.

We start by varying the interval length $t_c$. 
We display in Fig.~\ref{fig:SM_Sweep_tc}(a) the spectrum of the first (blue) and the last (red) singular value (SV) of $H$ as a function of $t_c$. The time axis is normalized by the maximum time which is inversely proportional to the frequency step: $t_{\mathrm{max}} = 1/\delta\nu$. These states correspond respectively to the maximum and minimum possible reflection within the output interval $\Delta t_y$. The first SV is zero for $t_c=0$, as expected, and then exponentially increases to rapidly reach a plateau at $\lambda_1 \approx 0.84$ shortly after $t_c=t_{c2}=0.05 t_{\mathrm{max}}$.
Thus, for $t>t_{c2}$, the interval length does not impact the reflection coefficient.  
On the other hand, the last SV is minimum for $t_{c2}<t<0.8t_{\mathrm{max}}$ with an average value $\langle\lambda_{N_\Lambda}\rangle \approx 2\times 10^{-11}$. For very short and  large values of $t_c$ and $t_c$, it is however not possible to shape the incident wavefront temporally to modulate the reflection coefficient. This implies that VPA states presented in the main text do not exist when $t_c$ is close to $0$ or $t_{\mathrm{max}}$.

The temporal responses for four time intervals  ($t_{c1}=0.004t_{\mathrm{max}}$, $t_{c2}=0.05t_{\mathrm{max}}$, $t_{c3}=0.35t_{\mathrm{max}}$ and $t_{c4} = 0.9t_{\mathrm{max}}$) are shown in the four last rows of Fig.~\ref{fig:SM_Sweep_tc}. These signals are reconstructed numerically from measurements of the scattering matrix. On the first column,  input (blue) and output (red) signals upon injecting the state corresponding to the maximum possible amount of energy are shown for the four time intervals considered. RTE's input and output signals, i.e. signals corresponding to minimal reflection, are displayed in the same way on the second column. In order to better observe the signals on the first line ($t_c=0.004 t_{\mathrm{max}}$), the time axis is adjusted to $[0 \ 0.03]\mu s$.

\section{Number of reflectionless transient states}
As explained in the main text, the number of reflectionless states found from the spatio-temporal operator $H$ is theoretically equal to the number of zeros located in the upper complex plane. The total number of zero within the frequency range $[f_1 = \omega_1/(2\pi);f_2 = \omega_2/(2\pi)]$ is equal to the number of poles of the scattering matrix and can be estimated from Weyl's law : $N_\omega = \int_{\omega_1}^{\omega_2} \rho(\omega)d\omega$, where $\rho(\omega) \sim A\omega / (2 \pi c_0^2)$ is the density of states. Here $A$ is the area of the cavity.

In this section, we provide further evidence of the relation between the number of reflectionless transient states and the number of zeros with positive imaginary parts. To do so, we vary the number of channels coupled to the cavity, from $N=2$ (corresponding to measurements presented in the main text) to $N=8$. In each case, we measure the $N\times N$ scattering matrix and obtain $\tau_W(\nu)$.  

\begin{figure}
    \centering
    \includegraphics[width=8.5cm]{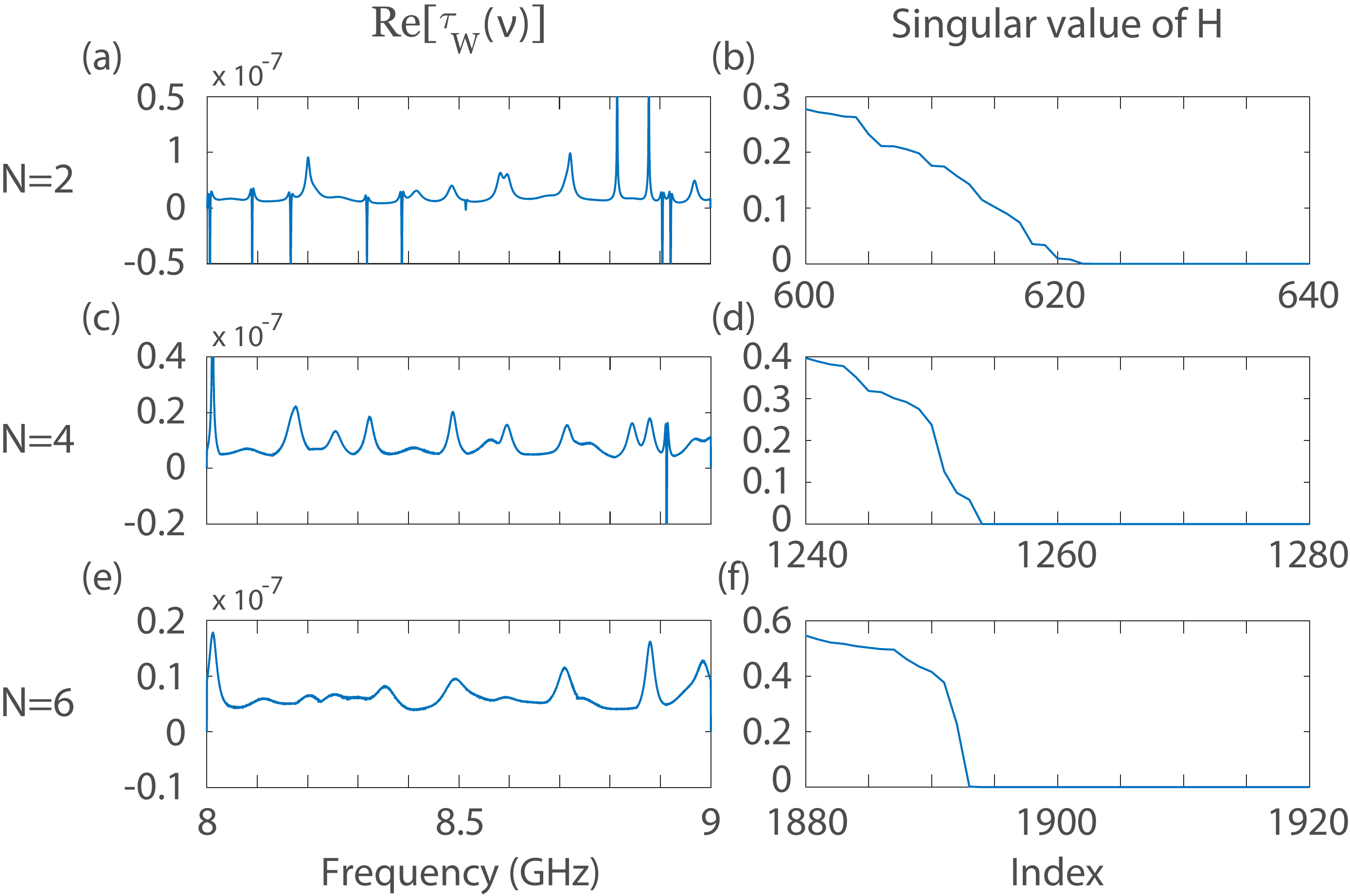}
    \caption{(a,c,e) Spectra of the real part of the trace of the Wigner-Smith operator $Q(\nu)=-i \left[ S(\nu) \right]^{-1}  [\partial S(\nu)/ \partial \nu]$ and (b,d,f) spectra of last singular values of $H$ for $N=2$ (a,b), $N=4$ (c,d) and $N=6$ (e,f).}
    \label{fig:SM_SV_WSO}
\end{figure}

The real part of the Wigner time delay is represented on the first column of  Fig.~\ref{fig:SM_SV_WSO} for different number of coupled channels $N$, $N=2$ (a), $N=4$ (c) and $N=6$ (e). For each case, the corresponding spectrum of the last $40$ singular values of $H$ is shown on the second column. 

For $N=2$, the Wigner time delay comprises positive and negative peaks. We find $18$ singular values of $H$ bellow $\lambda = 10^{-5}$. The small difference from the number presented in the main text is due to the slight changes in the system. This time, all the channels are matched with coax-to-waveguide transitions. For each measurements, $N$ ports are connected to the Vector Network Analyser, the others being in open-circuit states. This pushes the reflective boundaries away from the center of the cavity and so virtually increases its size.
For $N=4$, the Wigner time delay comprises only one negative peak and $25$ singular values below $\lambda = 10^{-5}$ are found.
For $N=6$, the peaks of $\tau_W(\nu)$ are all positive so that we expect that the number of reflectionless states to be equal to the number of resonances (or poles) with the bandwidth. We find $25$ singular values of $H$ smaller than $\lambda=10^{-5}$ which is in good agreement with Weyl's estimate.

\bibliographystyle{apsrev4-1}

\begin{thebibliography}{76}%
\makeatletter
\providecommand \@ifxundefined [1]{%
 \@ifx{#1\undefined}
}%
\providecommand \@ifnum [1]{%
 \ifnum #1\expandafter \@firstoftwo
 \else \expandafter \@secondoftwo
 \fi
}%
\providecommand \@ifx [1]{%
 \ifx #1\expandafter \@firstoftwo
 \else \expandafter \@secondoftwo
 \fi
}%
\providecommand \natexlab [1]{#1}%
\providecommand \enquote  [1]{``#1''}%
\providecommand \bibnamefont  [1]{#1}%
\providecommand \bibfnamefont [1]{#1}%
\providecommand \citenamefont [1]{#1}%
\providecommand \href@noop [0]{\@secondoftwo}%
\providecommand \href [0]{\begingroup \@sanitize@url \@href}%
\providecommand \@href[1]{\@@startlink{#1}\@@href}%
\providecommand \@@href[1]{\endgroup#1\@@endlink}%
\providecommand \@sanitize@url [0]{\catcode `\\12\catcode `\$12\catcode
  `\&12\catcode `\#12\catcode `\^12\catcode `\_12\catcode `\%12\relax}%
\providecommand \@@startlink[1]{}%
\providecommand \@@endlink[0]{}%
\providecommand \url  [0]{\begingroup\@sanitize@url \@url }%
\providecommand \@url [1]{\endgroup\@href {#1}{\urlprefix }}%
\providecommand \urlprefix  [0]{URL }%
\providecommand \Eprint [0]{\href }%
\providecommand \doibase [0]{http://dx.doi.org/}%
\providecommand \selectlanguage [0]{\@gobble}%
\providecommand \bibinfo  [0]{\@secondoftwo}%
\providecommand \bibfield  [0]{\@secondoftwo}%
\providecommand \translation [1]{[#1]}%
\providecommand \BibitemOpen [0]{}%
\providecommand \bibitemStop [0]{}%
\providecommand \bibitemNoStop [0]{.\EOS\space}%
\providecommand \EOS [0]{\spacefactor3000\relax}%
\providecommand \BibitemShut  [1]{\csname bibitem#1\endcsname}%
\let\auto@bib@innerbib\@empty
\bibitem [{\citenamefont {Rotter}\ and\ \citenamefont
  {Gigan}(2017)}]{rotter2017light}%
  \BibitemOpen
  \bibfield  {author} {\bibinfo {author} {\bibfnamefont {S.}~\bibnamefont
  {Rotter}}\ and\ \bibinfo {author} {\bibfnamefont {S.}~\bibnamefont {Gigan}},\
  }\href@noop {} {\bibfield  {journal} {\bibinfo  {journal} {Rev. Mod. Phys.}\
  }\textbf {\bibinfo {volume} {89}},\ \bibinfo {pages} {015005} (\bibinfo
  {year} {2017})}\BibitemShut {NoStop}%
\bibitem [{\citenamefont {Cao}\ \emph {et~al.}(2022)\citenamefont {Cao},
  \citenamefont {Mosk},\ and\ \citenamefont {Rotter}}]{Cao2022}%
  \BibitemOpen
  \bibfield  {author} {\bibinfo {author} {\bibfnamefont {H.}~\bibnamefont
  {Cao}}, \bibinfo {author} {\bibfnamefont {A.~P.}\ \bibnamefont {Mosk}}, \
  and\ \bibinfo {author} {\bibfnamefont {S.}~\bibnamefont {Rotter}},\
  }\href@noop {} {\bibfield  {journal} {\bibinfo  {journal} {Nature Physics}\
  }\textbf {\bibinfo {volume} {18}},\ \bibinfo {pages} {994} (\bibinfo {year}
  {2022})}\BibitemShut {NoStop}%
\bibitem [{\citenamefont {Vellekoop}\ and\ \citenamefont
  {Mosk}(2007)}]{Vellekoop2007}%
  \BibitemOpen
  \bibfield  {author} {\bibinfo {author} {\bibfnamefont {I.~M.}\ \bibnamefont
  {Vellekoop}}\ and\ \bibinfo {author} {\bibfnamefont {A.~P.}\ \bibnamefont
  {Mosk}},\ }\href@noop {} {\bibfield  {journal} {\bibinfo  {journal} {Opt.
  Lett.}\ }\textbf {\bibinfo {volume} {32}},\ \bibinfo {pages} {2309} (\bibinfo
  {year} {2007})}\BibitemShut {NoStop}%
\bibitem [{\citenamefont {Vellekoop}\ \emph {et~al.}(2010)\citenamefont
  {Vellekoop}, \citenamefont {Lagendijk},\ and\ \citenamefont
  {Mosk}}]{Vellekoop2010}%
  \BibitemOpen
  \bibfield  {author} {\bibinfo {author} {\bibfnamefont {I.~M.}\ \bibnamefont
  {Vellekoop}}, \bibinfo {author} {\bibfnamefont {A.}~\bibnamefont
  {Lagendijk}}, \ and\ \bibinfo {author} {\bibfnamefont {A.~P.}\ \bibnamefont
  {Mosk}},\ }\href@noop {} {\bibfield  {journal} {\bibinfo  {journal} {Nature
  Photon.}\ }\textbf {\bibinfo {volume} {4}},\ \bibinfo {pages} {320} (\bibinfo
  {year} {2010})},\ \bibinfo {note} {10.1038/nphoton.2010.3}\BibitemShut
  {NoStop}%
\bibitem [{\citenamefont {Ma}\ \emph {et~al.}(2014)\citenamefont {Ma},
  \citenamefont {Xu}, \citenamefont {Liu},\ and\ \citenamefont
  {Wang}}]{Ma2014}%
  \BibitemOpen
  \bibfield  {author} {\bibinfo {author} {\bibfnamefont {C.}~\bibnamefont
  {Ma}}, \bibinfo {author} {\bibfnamefont {X.}~\bibnamefont {Xu}}, \bibinfo
  {author} {\bibfnamefont {Y.}~\bibnamefont {Liu}}, \ and\ \bibinfo {author}
  {\bibfnamefont {L.~V.}\ \bibnamefont {Wang}},\ }\href@noop {} {\bibfield
  {journal} {\bibinfo  {journal} {Nature Photonics}\ }\textbf {\bibinfo
  {volume} {8}},\ \bibinfo {pages} {931} (\bibinfo {year} {2014})}\BibitemShut
  {NoStop}%
\bibitem [{\citenamefont {Horstmeyer}\ \emph {et~al.}(2015)\citenamefont
  {Horstmeyer}, \citenamefont {Ruan},\ and\ \citenamefont
  {Yang}}]{Horstmeyer2015}%
  \BibitemOpen
  \bibfield  {author} {\bibinfo {author} {\bibfnamefont {R.}~\bibnamefont
  {Horstmeyer}}, \bibinfo {author} {\bibfnamefont {H.}~\bibnamefont {Ruan}}, \
  and\ \bibinfo {author} {\bibfnamefont {C.}~\bibnamefont {Yang}},\ }\href@noop
  {} {\bibfield  {journal} {\bibinfo  {journal} {Nat Photon}\ }\textbf
  {\bibinfo {volume} {9}},\ \bibinfo {pages} {563} (\bibinfo {year}
  {2015})}\BibitemShut {NoStop}%
\bibitem [{\citenamefont {Tanter}\ \emph {et~al.}(2000)\citenamefont {Tanter},
  \citenamefont {Thomas},\ and\ \citenamefont {Fink}}]{Tanter2000}%
  \BibitemOpen
  \bibfield  {author} {\bibinfo {author} {\bibfnamefont {M.}~\bibnamefont
  {Tanter}}, \bibinfo {author} {\bibfnamefont {J.-L.}\ \bibnamefont {Thomas}},
  \ and\ \bibinfo {author} {\bibfnamefont {M.}~\bibnamefont {Fink}},\
  }\href@noop {} {\bibfield  {journal} {\bibinfo  {journal} {The Journal of the
  Acoustical Society of America}\ }\textbf {\bibinfo {volume} {108}},\ \bibinfo
  {pages} {223} (\bibinfo {year} {2000})}\BibitemShut {NoStop}%
\bibitem [{\citenamefont {Vellekoop}\ and\ \citenamefont
  {Mosk}(2008)}]{Vellekoop2008a}%
  \BibitemOpen
  \bibfield  {author} {\bibinfo {author} {\bibfnamefont {I.~M.}\ \bibnamefont
  {Vellekoop}}\ and\ \bibinfo {author} {\bibfnamefont {A.~P.}\ \bibnamefont
  {Mosk}},\ }\href@noop {} {\bibfield  {journal} {\bibinfo  {journal} {Phys.
  Rev. Lett.}\ }\textbf {\bibinfo {volume} {101}},\ \bibinfo {pages} {120601}
  (\bibinfo {year} {2008})}\BibitemShut {NoStop}%
\bibitem [{\citenamefont {Popoff}\ \emph {et~al.}(2010)\citenamefont {Popoff},
  \citenamefont {Lerosey}, \citenamefont {Carminati}, \citenamefont {Fink},
  \citenamefont {Boccara},\ and\ \citenamefont {Gigan}}]{Popoff2010}%
  \BibitemOpen
  \bibfield  {author} {\bibinfo {author} {\bibfnamefont {S.~M.}\ \bibnamefont
  {Popoff}}, \bibinfo {author} {\bibfnamefont {G.}~\bibnamefont {Lerosey}},
  \bibinfo {author} {\bibfnamefont {R.}~\bibnamefont {Carminati}}, \bibinfo
  {author} {\bibfnamefont {M.}~\bibnamefont {Fink}}, \bibinfo {author}
  {\bibfnamefont {A.~C.}\ \bibnamefont {Boccara}}, \ and\ \bibinfo {author}
  {\bibfnamefont {S.}~\bibnamefont {Gigan}},\ }\href@noop {} {\bibfield
  {journal} {\bibinfo  {journal} {Phys. Rev. Lett.}\ }\textbf {\bibinfo
  {volume} {104}},\ \bibinfo {pages} {100601} (\bibinfo {year}
  {2010})}\BibitemShut {NoStop}%
\bibitem [{\citenamefont {Popoff}\ \emph {et~al.}(2011)\citenamefont {Popoff},
  \citenamefont {Aubry}, \citenamefont {Lerosey}, \citenamefont {Fink},
  \citenamefont {Boccara},\ and\ \citenamefont {Gigan}}]{popoff2011exploiting}%
  \BibitemOpen
  \bibfield  {author} {\bibinfo {author} {\bibfnamefont {S.~M.}\ \bibnamefont
  {Popoff}}, \bibinfo {author} {\bibfnamefont {A.}~\bibnamefont {Aubry}},
  \bibinfo {author} {\bibfnamefont {G.}~\bibnamefont {Lerosey}}, \bibinfo
  {author} {\bibfnamefont {M.}~\bibnamefont {Fink}}, \bibinfo {author}
  {\bibfnamefont {A.-C.}\ \bibnamefont {Boccara}}, \ and\ \bibinfo {author}
  {\bibfnamefont {S.}~\bibnamefont {Gigan}},\ }\href@noop {} {\bibfield
  {journal} {\bibinfo  {journal} {Phys. Rev. Lett.}\ }\textbf {\bibinfo
  {volume} {107}},\ \bibinfo {pages} {263901} (\bibinfo {year}
  {2011})}\BibitemShut {NoStop}%
\bibitem [{\citenamefont {Chong}\ \emph {et~al.}(2010)\citenamefont {Chong},
  \citenamefont {Ge}, \citenamefont {Cao},\ and\ \citenamefont
  {Stone}}]{chong2010coherent}%
  \BibitemOpen
  \bibfield  {author} {\bibinfo {author} {\bibfnamefont {Y.D.}~\bibnamefont
  {Chong}}, \bibinfo {author} {\bibfnamefont {L.}~\bibnamefont {Ge}}, \bibinfo
  {author} {\bibfnamefont {H.}~\bibnamefont {Cao}}, \ and\ \bibinfo {author}
  {\bibfnamefont {A.~D.}\ \bibnamefont {Stone}},\ }\href@noop {} {\bibfield
  {journal} {\bibinfo  {journal} {Phys. Rev. Lett.}\ }\textbf {\bibinfo
  {volume} {105}},\ \bibinfo {pages} {053901} (\bibinfo {year}
  {2010})}\BibitemShut {NoStop}%
\bibitem [{\citenamefont {Wan}\ \emph {et~al.}(2011)\citenamefont {Wan},
  \citenamefont {Chong}, \citenamefont {Ge}, \citenamefont {Noh}, \citenamefont
  {Stone},\ and\ \citenamefont {Cao}}]{wan2011time}%
  \BibitemOpen
  \bibfield  {author} {\bibinfo {author} {\bibfnamefont {W.}~\bibnamefont
  {Wan}}, \bibinfo {author} {\bibfnamefont {Y.}~\bibnamefont {Chong}}, \bibinfo
  {author} {\bibfnamefont {L.}~\bibnamefont {Ge}}, \bibinfo {author}
  {\bibfnamefont {H.}~\bibnamefont {Noh}}, \bibinfo {author} {\bibfnamefont
  {A.~D.}\ \bibnamefont {Stone}}, \ and\ \bibinfo {author} {\bibfnamefont
  {H.}~\bibnamefont {Cao}},\ }\href@noop {} {\bibfield  {journal} {\bibinfo
  {journal} {Science}\ }\textbf {\bibinfo {volume} {331}},\ \bibinfo {pages}
  {889} (\bibinfo {year} {2011})}\BibitemShut {NoStop}%
\bibitem [{\citenamefont {Devaud}\ \emph {et~al.}(2021)\citenamefont {Devaud},
  \citenamefont {Rauer}, \citenamefont {Melchard}, \citenamefont {K{\"u}hmayer},
  \citenamefont {Rotter},\ and\ \citenamefont {Gigan}}]{Devaud2021}%
  \BibitemOpen
  \bibfield  {author} {\bibinfo {author} {\bibfnamefont {L.}~\bibnamefont
  {Devaud}}, \bibinfo {author} {\bibfnamefont {B.}~\bibnamefont {Rauer}},
  \bibinfo {author} {\bibfnamefont {J.}~\bibnamefont {Melchard}}, \bibinfo
  {author} {\bibfnamefont {M.}~\bibnamefont {K{\"u}hmayer}}, \bibinfo {author}
  {\bibfnamefont {S.}~\bibnamefont {Rotter}}, \ and\ \bibinfo {author}
  {\bibfnamefont {S.}~\bibnamefont {Gigan}},\ }\href@noop {} {\bibfield
  {journal} {\bibinfo  {journal} {Phys. Rev. Lett.}\ }\textbf {\bibinfo
  {volume} {127}},\ \bibinfo {pages} {093903} (\bibinfo {year}
  {2021})}\BibitemShut {NoStop}%
\bibitem [{\citenamefont {Kim}\ \emph {et~al.}(2012)\citenamefont {Kim},
  \citenamefont {Choi}, \citenamefont {Yoon}, \citenamefont {Choi},
  \citenamefont {Kim}, \citenamefont {Park},\ and\ \citenamefont
  {Choi}}]{kim2012maximal}%
  \BibitemOpen
  \bibfield  {author} {\bibinfo {author} {\bibfnamefont {M.}~\bibnamefont
  {Kim}}, \bibinfo {author} {\bibfnamefont {Y.}~\bibnamefont {Choi}}, \bibinfo
  {author} {\bibfnamefont {C.}~\bibnamefont {Yoon}}, \bibinfo {author}
  {\bibfnamefont {W.}~\bibnamefont {Choi}}, \bibinfo {author} {\bibfnamefont
  {J.}~\bibnamefont {Kim}}, \bibinfo {author} {\bibfnamefont {Q.-H.}\
  \bibnamefont {Park}}, \ and\ \bibinfo {author} {\bibfnamefont
  {W.}~\bibnamefont {Choi}},\ }\href@noop {} {\bibfield  {journal} {\bibinfo
  {journal} {Nat. Photonics}\ }\textbf {\bibinfo {volume} {6}},\ \bibinfo
  {pages} {581} (\bibinfo {year} {2012})}\BibitemShut {NoStop}%
\bibitem [{\citenamefont {G{\'e}rardin}\ \emph {et~al.}(2014)\citenamefont
  {G{\'e}rardin}, \citenamefont {Laurent}, \citenamefont {Derode},
  \citenamefont {Prada},\ and\ \citenamefont {Aubry}}]{Gerardin2014}%
  \BibitemOpen
  \bibfield  {author} {\bibinfo {author} {\bibfnamefont {B.}~\bibnamefont
  {G{\'e}rardin}}, \bibinfo {author} {\bibfnamefont {J.}~\bibnamefont
  {Laurent}}, \bibinfo {author} {\bibfnamefont {A.}~\bibnamefont {Derode}},
  \bibinfo {author} {\bibfnamefont {C.}~\bibnamefont {Prada}}, \ and\ \bibinfo
  {author} {\bibfnamefont {A.}~\bibnamefont {Aubry}},\ }\href@noop {}
  {\bibfield  {journal} {\bibinfo  {journal} {Phys. Rev. Lett.}\ }\textbf
  {\bibinfo {volume} {113}},\ \bibinfo {pages} {173901} (\bibinfo {year}
  {2014})}\BibitemShut {NoStop}%
\bibitem [{\citenamefont {Sarma}\ \emph {et~al.}(2016)\citenamefont {Sarma},
  \citenamefont {Yamilov}, \citenamefont {Petrenko}, \citenamefont {Bromberg},\
  and\ \citenamefont {Cao}}]{Sarma2016}%
  \BibitemOpen
  \bibfield  {author} {\bibinfo {author} {\bibfnamefont {R.}~\bibnamefont
  {Sarma}}, \bibinfo {author} {\bibfnamefont {A.~G.}\ \bibnamefont {Yamilov}},
  \bibinfo {author} {\bibfnamefont {S.}~\bibnamefont {Petrenko}}, \bibinfo
  {author} {\bibfnamefont {Y.}~\bibnamefont {Bromberg}}, \ and\ \bibinfo
  {author} {\bibfnamefont {H.}~\bibnamefont {Cao}},\ }\href@noop {} {\bibfield
  {journal} {\bibinfo  {journal} {Phys. Rev. Lett.}\ }\textbf {\bibinfo
  {volume} {117}},\ \bibinfo {pages} {086803} (\bibinfo {year}
  {2016})}\BibitemShut {NoStop}%
\bibitem [{\citenamefont {Jeong}\ \emph {et~al.}(2018)\citenamefont {Jeong},
  \citenamefont {Lee}, \citenamefont {Choi}, \citenamefont {Kang},
  \citenamefont {Hong}, \citenamefont {Park}, \citenamefont {Lim},
  \citenamefont {Park},\ and\ \citenamefont {Choi}}]{Jeong18}%
  \BibitemOpen
  \bibfield  {author} {\bibinfo {author} {\bibfnamefont {S.}~\bibnamefont
  {Jeong}}, \bibinfo {author} {\bibfnamefont {Y.-R.}\ \bibnamefont {Lee}},
  \bibinfo {author} {\bibfnamefont {W.}~\bibnamefont {Choi}}, \bibinfo {author}
  {\bibfnamefont {S.}~\bibnamefont {Kang}}, \bibinfo {author} {\bibfnamefont
  {J.~H.}\ \bibnamefont {Hong}}, \bibinfo {author} {\bibfnamefont {J.-S.}\
  \bibnamefont {Park}}, \bibinfo {author} {\bibfnamefont {Y.-S.}\ \bibnamefont
  {Lim}}, \bibinfo {author} {\bibfnamefont {H.-G.}\ \bibnamefont {Park}}, \
  and\ \bibinfo {author} {\bibfnamefont {W.}~\bibnamefont {Choi}},\ }\href@noop
  {} {\bibfield  {journal} {\bibinfo  {journal} {Nat. Photonics}\ }\textbf
  {\bibinfo {volume} {12}},\ \bibinfo {pages} {277} (\bibinfo {year}
  {2018})}\BibitemShut {NoStop}%
\bibitem [{\citenamefont {Bender}\ \emph {et~al.}(2022)\citenamefont {Bender},
  \citenamefont {Yamilov}, \citenamefont {Goetschy}, \citenamefont {Yilmaz},
  \citenamefont {Hsu},\ and\ \citenamefont {Cao}}]{Bender2022}%
  \BibitemOpen
  \bibfield  {author} {\bibinfo {author} {\bibfnamefont {N.}~\bibnamefont
  {Bender}}, \bibinfo {author} {\bibfnamefont {A.}~\bibnamefont {Yamilov}},
  \bibinfo {author} {\bibfnamefont {A.}~\bibnamefont {Goetschy}}, \bibinfo
  {author} {\bibfnamefont {H.}~\bibnamefont {Yilmaz}}, \bibinfo {author}
  {\bibfnamefont {C.~W.}\ \bibnamefont {Hsu}}, \ and\ \bibinfo {author}
  {\bibfnamefont {H.}~\bibnamefont {Cao}},\ }\href@noop {} {\bibfield
  {journal} {\bibinfo  {journal} {Nat. Phys.}\ } (\bibinfo {year}
  {2022})}\BibitemShut {NoStop}%
\bibitem [{\citenamefont {Durand}\ \emph {et~al.}(2019)\citenamefont {Durand},
  \citenamefont {Popoff}, \citenamefont {Carminati},\ and\ \citenamefont
  {Goetschy}}]{Durand2019}%
  \BibitemOpen
  \bibfield  {author} {\bibinfo {author} {\bibfnamefont {M.}~\bibnamefont
  {Durand}}, \bibinfo {author} {\bibfnamefont {S.~M.}\ \bibnamefont {Popoff}},
  \bibinfo {author} {\bibfnamefont {R.}~\bibnamefont {Carminati}}, \ and\
  \bibinfo {author} {\bibfnamefont {A.}~\bibnamefont {Goetschy}},\ }\href@noop
  {} {\bibfield  {journal} {\bibinfo  {journal} {Phys. Rev. Lett.}\ }\textbf
  {\bibinfo {volume} {123}},\ \bibinfo {pages} {243901} (\bibinfo {year}
  {2019})}\BibitemShut {NoStop}%
\bibitem [{\citenamefont {del Hougne}\ \emph
  {et~al.}(2021{\natexlab{a}})\citenamefont {del Hougne}, \citenamefont
  {Sobry}, \citenamefont {Legrand}, \citenamefont {Mortessagne}, \citenamefont
  {Kuhl},\ and\ \citenamefont {Davy}}]{delHougne2021}%
  \BibitemOpen
  \bibfield  {author} {\bibinfo {author} {\bibfnamefont {P.}~\bibnamefont {del
  Hougne}}, \bibinfo {author} {\bibfnamefont {R.}~\bibnamefont {Sobry}},
  \bibinfo {author} {\bibfnamefont {O.}~\bibnamefont {Legrand}}, \bibinfo
  {author} {\bibfnamefont {F.}~\bibnamefont {Mortessagne}}, \bibinfo {author}
  {\bibfnamefont {U.}~\bibnamefont {Kuhl}}, \ and\ \bibinfo {author}
  {\bibfnamefont {M.}~\bibnamefont {Davy}},\ }\href@noop {} {\bibfield
  {journal} {\bibinfo  {journal} {Laser Photonics Rev.}\ ,\ \bibinfo {pages}
  {2000335}} (\bibinfo {year} {2021}{\natexlab{a}})}\BibitemShut {NoStop}%
\bibitem [{\citenamefont {Horodynski}\ \emph {et~al.}(2019)\citenamefont
  {Horodynski}, \citenamefont {K{\"u}hmayer}, \citenamefont {Brandstotter},
  \citenamefont {Pichler}, \citenamefont {Fyodorov}, \citenamefont {Kuhl},\
  and\ \citenamefont {Rotter}}]{Horodynski2019NatPhot}%
  \BibitemOpen
  \bibfield  {author} {\bibinfo {author} {\bibfnamefont {M.}~\bibnamefont
  {Horodynski}}, \bibinfo {author} {\bibfnamefont {M.}~\bibnamefont
  {K{\"u}hmayer}}, \bibinfo {author} {\bibfnamefont {A.}~\bibnamefont
  {Brandstotter}}, \bibinfo {author} {\bibfnamefont {K.}~\bibnamefont
  {Pichler}}, \bibinfo {author} {\bibfnamefont {Y.~V.}\ \bibnamefont
  {Fyodorov}}, \bibinfo {author} {\bibfnamefont {U.}~\bibnamefont {Kuhl}}, \
  and\ \bibinfo {author} {\bibfnamefont {S.}~\bibnamefont {Rotter}},\
  }\href@noop {} {\bibfield  {journal} {\bibinfo  {journal} {Nature Photon.}\
  ,\ \bibinfo {pages} {1}} (\bibinfo {year} {2019})}\BibitemShut {NoStop}%
\bibitem [{\citenamefont {Bouchet}\ \emph {et~al.}(2021)\citenamefont
  {Bouchet}, \citenamefont {Rotter},\ and\ \citenamefont
  {Mosk}}]{Bouchet2021NatPhys}%
  \BibitemOpen
  \bibfield  {author} {\bibinfo {author} {\bibfnamefont {D.}~\bibnamefont
  {Bouchet}}, \bibinfo {author} {\bibfnamefont {S.}~\bibnamefont {Rotter}}, \
  and\ \bibinfo {author} {\bibfnamefont {A.~P.}\ \bibnamefont {Mosk}},\
  }\href@noop {} {\bibfield  {journal} {\bibinfo  {journal} {Nat. Phys.}\
  }\textbf {\bibinfo {volume} {17}},\ \bibinfo {pages} {564} (\bibinfo {year}
  {2021})}\BibitemShut {NoStop}%
\bibitem [{\citenamefont {Aulbach}\ \emph {et~al.}(2011)\citenamefont
  {Aulbach}, \citenamefont {Gjonaj}, \citenamefont {Johnson}, \citenamefont
  {Mosk},\ and\ \citenamefont {Lagendijk}}]{Aulbach2011}%
  \BibitemOpen
  \bibfield  {author} {\bibinfo {author} {\bibfnamefont {J.}~\bibnamefont
  {Aulbach}}, \bibinfo {author} {\bibfnamefont {B.}~\bibnamefont {Gjonaj}},
  \bibinfo {author} {\bibfnamefont {P.~M.}\ \bibnamefont {Johnson}}, \bibinfo
  {author} {\bibfnamefont {A.~P.}\ \bibnamefont {Mosk}}, \ and\ \bibinfo
  {author} {\bibfnamefont {A.}~\bibnamefont {Lagendijk}},\ }\href@noop {}
  {\bibfield  {journal} {\bibinfo  {journal} {Phys. Rev. Lett.}\ }\textbf
  {\bibinfo {volume} {106}},\ \bibinfo {pages} {103901} (\bibinfo {year}
  {2011})}\BibitemShut {NoStop}%
\bibitem [{\citenamefont {Katz}\ \emph {et~al.}(2011)\citenamefont {Katz},
  \citenamefont {Small}, \citenamefont {Bromberg},\ and\ \citenamefont
  {Silberberg}}]{katz2011focusing}%
  \BibitemOpen
  \bibfield  {author} {\bibinfo {author} {\bibfnamefont {O.}~\bibnamefont
  {Katz}}, \bibinfo {author} {\bibfnamefont {E.}~\bibnamefont {Small}},
  \bibinfo {author} {\bibfnamefont {Y.}~\bibnamefont {Bromberg}}, \ and\
  \bibinfo {author} {\bibfnamefont {Y.}~\bibnamefont {Silberberg}},\
  }\href@noop {} {\bibfield  {journal} {\bibinfo  {journal} {Nat. Photonics}\
  }\textbf {\bibinfo {volume} {5}},\ \bibinfo {pages} {372} (\bibinfo {year}
  {2011})}\BibitemShut {NoStop}%
\bibitem [{\citenamefont {Andreoli}\ \emph {et~al.}(2015)\citenamefont
  {Andreoli}, \citenamefont {Volpe}, \citenamefont {Popoff}, \citenamefont
  {Katz}, \citenamefont {Gr{\'e}sillon},\ and\ \citenamefont
  {Gigan}}]{andreoli2015deterministic}%
  \BibitemOpen
  \bibfield  {author} {\bibinfo {author} {\bibfnamefont {D.}~\bibnamefont
  {Andreoli}}, \bibinfo {author} {\bibfnamefont {G.}~\bibnamefont {Volpe}},
  \bibinfo {author} {\bibfnamefont {S.}~\bibnamefont {Popoff}}, \bibinfo
  {author} {\bibfnamefont {O.}~\bibnamefont {Katz}}, \bibinfo {author}
  {\bibfnamefont {S.}~\bibnamefont {Gr{\'e}sillon}}, \ and\ \bibinfo {author}
  {\bibfnamefont {S.}~\bibnamefont {Gigan}},\ }\href@noop {} {\bibfield
  {journal} {\bibinfo  {journal} {Sci. Rep.}\ }\textbf {\bibinfo {volume}
  {5}},\ \bibinfo {pages} {1} (\bibinfo {year} {2015})}\BibitemShut {NoStop}%
\bibitem [{\citenamefont {Mounaix}\ \emph
  {et~al.}(2016{\natexlab{a}})\citenamefont {Mounaix}, \citenamefont
  {Andreoli}, \citenamefont {Defienne}, \citenamefont {Volpe}, \citenamefont
  {Katz}, \citenamefont {Gresillon},\ and\ \citenamefont
  {Gigan}}]{Mounaix2016a}%
  \BibitemOpen
  \bibfield  {author} {\bibinfo {author} {\bibfnamefont {M.}~\bibnamefont
  {Mounaix}}, \bibinfo {author} {\bibfnamefont {D.}~\bibnamefont {Andreoli}},
  \bibinfo {author} {\bibfnamefont {H.}~\bibnamefont {Defienne}}, \bibinfo
  {author} {\bibfnamefont {G.}~\bibnamefont {Volpe}}, \bibinfo {author}
  {\bibfnamefont {O.}~\bibnamefont {Katz}}, \bibinfo {author} {\bibfnamefont
  {S.}~\bibnamefont {Gresillon}}, \ and\ \bibinfo {author} {\bibfnamefont
  {S.}~\bibnamefont {Gigan}},\ }\href@noop {} {\bibfield  {journal} {\bibinfo
  {journal} {Phys. Rev. Lett.}\ }\textbf {\bibinfo {volume} {116}},\ \bibinfo
  {pages} {253901} (\bibinfo {year} {2016}{\natexlab{a}})}\BibitemShut
  {NoStop}%
\bibitem [{\citenamefont {Choi}\ \emph {et~al.}(2013)\citenamefont {Choi},
  \citenamefont {Hillman}, \citenamefont {Choi}, \citenamefont {Lue},
  \citenamefont {Dasari}, \citenamefont {So}, \citenamefont {Choi},\ and\
  \citenamefont {Yaqoob}}]{ChoiPRL2013}%
  \BibitemOpen
  \bibfield  {author} {\bibinfo {author} {\bibfnamefont {Y.}~\bibnamefont
  {Choi}}, \bibinfo {author} {\bibfnamefont {T.~R.}\ \bibnamefont {Hillman}},
  \bibinfo {author} {\bibfnamefont {W.}~\bibnamefont {Choi}}, \bibinfo {author}
  {\bibfnamefont {N.}~\bibnamefont {Lue}}, \bibinfo {author} {\bibfnamefont
  {R.~R.}\ \bibnamefont {Dasari}}, \bibinfo {author} {\bibfnamefont {P.~T.~C.}\
  \bibnamefont {So}}, \bibinfo {author} {\bibfnamefont {W.}~\bibnamefont
  {Choi}}, \ and\ \bibinfo {author} {\bibfnamefont {Z.}~\bibnamefont
  {Yaqoob}},\ }\href@noop {} {\bibfield  {journal} {\bibinfo  {journal} {Phys.
  Rev. Lett.}\ }\textbf {\bibinfo {volume} {111}},\ \bibinfo {pages} {243901}
  (\bibinfo {year} {2013})}\BibitemShut {NoStop}%
\bibitem [{\citenamefont {Mounaix}\ \emph
  {et~al.}(2016{\natexlab{b}})\citenamefont {Mounaix}, \citenamefont
  {Defienne},\ and\ \citenamefont {Gigan}}]{Mounaix2016}%
  \BibitemOpen
  \bibfield  {author} {\bibinfo {author} {\bibfnamefont {M.}~\bibnamefont
  {Mounaix}}, \bibinfo {author} {\bibfnamefont {H.}~\bibnamefont {Defienne}}, \
  and\ \bibinfo {author} {\bibfnamefont {S.}~\bibnamefont {Gigan}},\
  }\href@noop {} {\bibfield  {journal} {\bibinfo  {journal} {Phys. Rev. A}\
  }\textbf {\bibinfo {volume} {94}},\ \bibinfo {pages} {041802(R)} (\bibinfo
  {year} {2016}{\natexlab{b}})}\BibitemShut {NoStop}%
\bibitem [{\citenamefont {Devaud}\ \emph {et~al.}(2022)\citenamefont {Devaud},
  \citenamefont {Rauer}, \citenamefont {K{\"u}hmayer}, \citenamefont
  {Melchard}, \citenamefont {Mounaix}, \citenamefont {Rotter},\ and\
  \citenamefont {Gigan}}]{devaud2022temporal}%
  \BibitemOpen
  \bibfield  {author} {\bibinfo {author} {\bibfnamefont {L.}~\bibnamefont
  {Devaud}}, \bibinfo {author} {\bibfnamefont {B.}~\bibnamefont {Rauer}},
  \bibinfo {author} {\bibfnamefont {M.}~\bibnamefont {K{\"u}hmayer}}, \bibinfo
  {author} {\bibfnamefont {J.}~\bibnamefont {Melchard}}, \bibinfo {author}
  {\bibfnamefont {M.}~\bibnamefont {Mounaix}}, \bibinfo {author} {\bibfnamefont
  {S.}~\bibnamefont {Rotter}}, \ and\ \bibinfo {author} {\bibfnamefont
  {S.}~\bibnamefont {Gigan}},\ }\href@noop {} {\bibfield  {journal} {\bibinfo
  {journal} {Phys. Rev. A}\ }\textbf {\bibinfo {volume} {105}},\ \bibinfo
  {pages} {L051501} (\bibinfo {year} {2022})}\BibitemShut {NoStop}%
\bibitem [{\citenamefont {del Hougne}\ \emph {et~al.}(2016)\citenamefont {del
  Hougne}, \citenamefont {Lemoult}, \citenamefont {Fink},\ and\ \citenamefont
  {Lerosey}}]{del2016spatiotemporal}%
  \BibitemOpen
  \bibfield  {author} {\bibinfo {author} {\bibfnamefont {P.}~\bibnamefont {del
  Hougne}}, \bibinfo {author} {\bibfnamefont {F.}~\bibnamefont {Lemoult}},
  \bibinfo {author} {\bibfnamefont {M.}~\bibnamefont {Fink}}, \ and\ \bibinfo
  {author} {\bibfnamefont {G.}~\bibnamefont {Lerosey}},\ }\href@noop {}
  {\bibfield  {journal} {\bibinfo  {journal} {Phys. Rev. Lett.}\ }\textbf
  {\bibinfo {volume} {117}},\ \bibinfo {pages} {134302} (\bibinfo {year}
  {2016})}\BibitemShut {NoStop}%
\bibitem [{\citenamefont {F.~Imani}\ \emph {et~al.}(2022)\citenamefont
  {F.~Imani}, \citenamefont {Abadal},\ and\ \citenamefont {del
  Hougne}}]{imani2022metasurface}%
  \BibitemOpen
  \bibfield  {author} {\bibinfo {author} {\bibfnamefont {M.}~\bibnamefont
  {F.~Imani}}, \bibinfo {author} {\bibfnamefont {S.}~\bibnamefont {Abadal}}, \
  and\ \bibinfo {author} {\bibfnamefont {P.}~\bibnamefont {del Hougne}},\
  }\href@noop {} {\bibfield  {journal} {\bibinfo  {journal} {Adv. Sci.}\
  }\textbf {\bibinfo {volume} {9}},\ \bibinfo {pages} {2201458} (\bibinfo
  {year} {2022})}\BibitemShut {NoStop}%
\bibitem [{\citenamefont {Rabault}\ \emph {et~al.}(2023)\citenamefont
  {Rabault}, \citenamefont {Magoarou}, \citenamefont {Sol}, \citenamefont
  {Alexandropoulos}, \citenamefont {Shlezinger}, \citenamefont {Poor},\ and\
  \citenamefont {del Hougne}}]{rabault2023tacit}%
  \BibitemOpen
  \bibfield  {author} {\bibinfo {author} {\bibfnamefont {A.}~\bibnamefont
  {Rabault}}, \bibinfo {author} {\bibfnamefont {L.~L.}\ \bibnamefont
  {Magoarou}}, \bibinfo {author} {\bibfnamefont {J.}~\bibnamefont {Sol}},
  \bibinfo {author} {\bibfnamefont {G.~C.}\ \bibnamefont {Alexandropoulos}},
  \bibinfo {author} {\bibfnamefont {N.}~\bibnamefont {Shlezinger}}, \bibinfo
  {author} {\bibfnamefont {H.~V.}\ \bibnamefont {Poor}}, \ and\ \bibinfo
  {author} {\bibfnamefont {P.}~\bibnamefont {del Hougne}},\ }\href@noop {}
  {\bibfield  {journal} {\bibinfo  {journal} {arXiv:2302.04993}\ } (\bibinfo
  {year} {2023})}\BibitemShut {NoStop}%
\bibitem [{\citenamefont {Mosk}\ \emph {et~al.}(2012)\citenamefont {Mosk},
  \citenamefont {Lagendijk}, \citenamefont {Lerosey},\ and\ \citenamefont
  {Fink}}]{Mosk2012}%
  \BibitemOpen
  \bibfield  {author} {\bibinfo {author} {\bibfnamefont {A.~P.}\ \bibnamefont
  {Mosk}}, \bibinfo {author} {\bibfnamefont {A.}~\bibnamefont {Lagendijk}},
  \bibinfo {author} {\bibfnamefont {G.}~\bibnamefont {Lerosey}}, \ and\
  \bibinfo {author} {\bibfnamefont {M.}~\bibnamefont {Fink}},\ }\href@noop {}
  {\bibfield  {journal} {\bibinfo  {journal} {Nat. Photonics}\ }\textbf
  {\bibinfo {volume} {6}},\ \bibinfo {pages} {283} (\bibinfo {year}
  {2012})}\BibitemShut {NoStop}%
\bibitem [{\citenamefont {Fink}(1992)}]{fink1992time}%
  \BibitemOpen
  \bibfield  {author} {\bibinfo {author} {\bibfnamefont {M.}~\bibnamefont
  {Fink}},\ }\href@noop {} {\bibfield  {journal} {\bibinfo  {journal} {IEEE
  Trans. Ultrason. Ferroelectr. Freq. Control}\ }\textbf {\bibinfo {volume}
  {39}},\ \bibinfo {pages} {555} (\bibinfo {year} {1992})}\BibitemShut
  {NoStop}%
\bibitem [{\citenamefont {Derode}\ \emph {et~al.}(1995)\citenamefont {Derode},
  \citenamefont {Roux},\ and\ \citenamefont {Fink}}]{Derode1995}%
  \BibitemOpen
  \bibfield  {author} {\bibinfo {author} {\bibfnamefont {A.}~\bibnamefont
  {Derode}}, \bibinfo {author} {\bibfnamefont {P.}~\bibnamefont {Roux}}, \ and\
  \bibinfo {author} {\bibfnamefont {M.}~\bibnamefont {Fink}},\ }\href@noop {}
  {\bibfield  {journal} {\bibinfo  {journal} {Phys. Rev. Lett.}\ }\textbf
  {\bibinfo {volume} {75}},\ \bibinfo {pages} {4206} (\bibinfo {year}
  {1995})}\BibitemShut {NoStop}%
\bibitem [{\citenamefont {Lemoult}\ \emph {et~al.}(2009)\citenamefont
  {Lemoult}, \citenamefont {Lerosey}, \citenamefont {de~Rosny},\ and\
  \citenamefont {Fink}}]{lemoult2009manipulating}%
  \BibitemOpen
  \bibfield  {author} {\bibinfo {author} {\bibfnamefont {F.}~\bibnamefont
  {Lemoult}}, \bibinfo {author} {\bibfnamefont {G.}~\bibnamefont {Lerosey}},
  \bibinfo {author} {\bibfnamefont {J.}~\bibnamefont {de~Rosny}}, \ and\
  \bibinfo {author} {\bibfnamefont {M.}~\bibnamefont {Fink}},\ }\href@noop {}
  {\bibfield  {journal} {\bibinfo  {journal} {Phys. Rev. Lett.}\ }\textbf
  {\bibinfo {volume} {103}},\ \bibinfo {pages} {173902} (\bibinfo {year}
  {2009})}\BibitemShut {NoStop}%
\bibitem [{\citenamefont {Draeger}\ and\ \citenamefont
  {Fink}(1997)}]{draeger1997one}%
  \BibitemOpen
  \bibfield  {author} {\bibinfo {author} {\bibfnamefont {C.}~\bibnamefont
  {Draeger}}\ and\ \bibinfo {author} {\bibfnamefont {M.}~\bibnamefont {Fink}},\
  }\href@noop {} {\bibfield  {journal} {\bibinfo  {journal} {Phys. Rev. Lett.}\
  }\textbf {\bibinfo {volume} {79}},\ \bibinfo {pages} {407} (\bibinfo {year}
  {1997})}\BibitemShut {NoStop}%
\bibitem [{\citenamefont {Lerosey}\ \emph {et~al.}(2007)\citenamefont
  {Lerosey}, \citenamefont {de~Rosny}, \citenamefont {Tourin},\ and\
  \citenamefont {Fink}}]{Lerosey2007}%
  \BibitemOpen
  \bibfield  {author} {\bibinfo {author} {\bibfnamefont {G.}~\bibnamefont
  {Lerosey}}, \bibinfo {author} {\bibfnamefont {J.}~\bibnamefont {de~Rosny}},
  \bibinfo {author} {\bibfnamefont {A.}~\bibnamefont {Tourin}}, \ and\ \bibinfo
  {author} {\bibfnamefont {M.}~\bibnamefont {Fink}},\ }\href@noop {} {\bibfield
   {journal} {\bibinfo  {journal} {Science}\ }\textbf {\bibinfo {volume}
  {315}},\ \bibinfo {pages} {1120} (\bibinfo {year} {2007})}\BibitemShut
  {NoStop}%
\bibitem [{\citenamefont {Bouchet}\ and\ \citenamefont
  {Bossy}(2022)}]{Bouchet2022}%
  \BibitemOpen
  \bibfield  {author} {\bibinfo {author} {\bibfnamefont {D.}~\bibnamefont
  {Bouchet}}\ and\ \bibinfo {author} {\bibfnamefont {E.}~\bibnamefont
  {Bossy}},\ }\href@noop {} {\bibfield  {journal} {\bibinfo  {journal} {arXiv
  preprint arXiv:2211.16903}\ } (\bibinfo {year} {2022})}\BibitemShut {NoStop}%
\bibitem [{\citenamefont {McCabe}\ \emph {et~al.}(2011)\citenamefont {McCabe},
  \citenamefont {Tajalli}, \citenamefont {Austin}, \citenamefont {Bondareff},
  \citenamefont {Walmsley}, \citenamefont {Gigan},\ and\ \citenamefont
  {Chatel}}]{McCabe2011}%
  \BibitemOpen
  \bibfield  {author} {\bibinfo {author} {\bibfnamefont {D.~J.}\ \bibnamefont
  {McCabe}}, \bibinfo {author} {\bibfnamefont {A.}~\bibnamefont {Tajalli}},
  \bibinfo {author} {\bibfnamefont {D.~R.}\ \bibnamefont {Austin}}, \bibinfo
  {author} {\bibfnamefont {P.}~\bibnamefont {Bondareff}}, \bibinfo {author}
  {\bibfnamefont {I.~A.}\ \bibnamefont {Walmsley}}, \bibinfo {author}
  {\bibfnamefont {S.}~\bibnamefont {Gigan}}, \ and\ \bibinfo {author}
  {\bibfnamefont {B.}~\bibnamefont {Chatel}},\ }\href@noop {} {\bibfield
  {journal} {\bibinfo  {journal} {Nat. Commun.}\ }\textbf {\bibinfo {volume}
  {2}},\ \bibinfo {pages} {447} (\bibinfo {year} {2011})}\BibitemShut {NoStop}%
\bibitem [{\citenamefont {Mounaix}\ \emph {et~al.}(2020)\citenamefont
  {Mounaix}, \citenamefont {Fontaine}, \citenamefont {Neilson}, \citenamefont
  {Ryf}, \citenamefont {Chen}, \citenamefont {Alvarado-Zacarias},\ and\
  \citenamefont {Carpenter}}]{mounaix2020time}%
  \BibitemOpen
  \bibfield  {author} {\bibinfo {author} {\bibfnamefont {M.}~\bibnamefont
  {Mounaix}}, \bibinfo {author} {\bibfnamefont {N.~K.}\ \bibnamefont
  {Fontaine}}, \bibinfo {author} {\bibfnamefont {D.~T.}\ \bibnamefont
  {Neilson}}, \bibinfo {author} {\bibfnamefont {R.}~\bibnamefont {Ryf}},
  \bibinfo {author} {\bibfnamefont {H.}~\bibnamefont {Chen}}, \bibinfo {author}
  {\bibfnamefont {J.~C.}\ \bibnamefont {Alvarado-Zacarias}}, \ and\ \bibinfo
  {author} {\bibfnamefont {J.}~\bibnamefont {Carpenter}},\ }\href@noop {}
  {\bibfield  {journal} {\bibinfo  {journal} {Nat. Commun.}\ }\textbf {\bibinfo
  {volume} {11}},\ \bibinfo {pages} {1} (\bibinfo {year} {2020})}\BibitemShut
  {NoStop}%
\bibitem [{\citenamefont {Baranov}\ \emph {et~al.}(2017)\citenamefont
  {Baranov}, \citenamefont {Krasnok},\ and\ \citenamefont
  {Alu}}]{Baranov2017optica}%
  \BibitemOpen
  \bibfield  {author} {\bibinfo {author} {\bibfnamefont {D.~G.}\ \bibnamefont
  {Baranov}}, \bibinfo {author} {\bibfnamefont {A.}~\bibnamefont {Krasnok}}, \
  and\ \bibinfo {author} {\bibfnamefont {A.}~\bibnamefont {Alu}},\ }\href@noop
  {} {\bibfield  {journal} {\bibinfo  {journal} {Optica}\ }\textbf {\bibinfo
  {volume} {4}},\ \bibinfo {pages} {1457} (\bibinfo {year} {2017})}\BibitemShut
  {NoStop}%
\bibitem [{\citenamefont {Pai}\ \emph {et~al.}(2021)\citenamefont {Pai},
  \citenamefont {Bosch}, \citenamefont {K{\"u}hmayer}, \citenamefont {Rotter},\
  and\ \citenamefont {Mosk}}]{Pai2021}%
  \BibitemOpen
  \bibfield  {author} {\bibinfo {author} {\bibfnamefont {P.}~\bibnamefont
  {Pai}}, \bibinfo {author} {\bibfnamefont {J.}~\bibnamefont {Bosch}}, \bibinfo
  {author} {\bibfnamefont {M.}~\bibnamefont {K{\"u}hmayer}}, \bibinfo {author}
  {\bibfnamefont {S.}~\bibnamefont {Rotter}}, \ and\ \bibinfo {author}
  {\bibfnamefont {A.~P.}\ \bibnamefont {Mosk}},\ }\href@noop {} {\bibfield
  {journal} {\bibinfo  {journal} {Nat. Photonics}\ }\textbf {\bibinfo {volume}
  {15}},\ \bibinfo {pages} {431} (\bibinfo {year} {2021})}\BibitemShut
  {NoStop}%
\bibitem [{\citenamefont {Trainiti}\ \emph {et~al.}(2019)\citenamefont
  {Trainiti}, \citenamefont {Ra'di}, \citenamefont {Ruzzene},\ and\
  \citenamefont {Alu}}]{Trainiti2019}%
  \BibitemOpen
  \bibfield  {author} {\bibinfo {author} {\bibfnamefont {G.}~\bibnamefont
  {Trainiti}}, \bibinfo {author} {\bibfnamefont {Y.}~\bibnamefont {Ra'di}},
  \bibinfo {author} {\bibfnamefont {M.}~\bibnamefont {Ruzzene}}, \ and\
  \bibinfo {author} {\bibfnamefont {A.}~\bibnamefont {Alu}},\ }\href@noop {}
  {\bibfield  {journal} {\bibinfo  {journal} {Sci. Adv.}\ }\textbf {\bibinfo
  {volume} {5}},\ \bibinfo {pages} {eaaw3255} (\bibinfo {year}
  {2019})}\BibitemShut {NoStop}%
\bibitem [{\citenamefont {Lepeshov}\ and\ \citenamefont
  {Krasnok}(2020)}]{Lepeshov2020}%
  \BibitemOpen
  \bibfield  {author} {\bibinfo {author} {\bibfnamefont {S.}~\bibnamefont
  {Lepeshov}}\ and\ \bibinfo {author} {\bibfnamefont {A.}~\bibnamefont
  {Krasnok}},\ }\href@noop {} {\bibfield  {journal} {\bibinfo  {journal}
  {Optica}\ }\textbf {\bibinfo {volume} {7}},\ \bibinfo {pages} {1024}
  (\bibinfo {year} {2020})}\BibitemShut {NoStop}%
\bibitem [{\citenamefont {Grigoriev}\ \emph
  {et~al.}(2013{\natexlab{a}})\citenamefont {Grigoriev}, \citenamefont {Tahri},
  \citenamefont {Varault}, \citenamefont {Rolly}, \citenamefont {Stout},
  \citenamefont {Wenger},\ and\ \citenamefont
  {Bonod}}]{grigoriev2013optimization}%
  \BibitemOpen
  \bibfield  {author} {\bibinfo {author} {\bibfnamefont {V.}~\bibnamefont
  {Grigoriev}}, \bibinfo {author} {\bibfnamefont {A.}~\bibnamefont {Tahri}},
  \bibinfo {author} {\bibfnamefont {S.}~\bibnamefont {Varault}}, \bibinfo
  {author} {\bibfnamefont {B.}~\bibnamefont {Rolly}}, \bibinfo {author}
  {\bibfnamefont {B.}~\bibnamefont {Stout}}, \bibinfo {author} {\bibfnamefont
  {J.}~\bibnamefont {Wenger}}, \ and\ \bibinfo {author} {\bibfnamefont
  {N.}~\bibnamefont {Bonod}},\ }\href@noop {} {\bibfield  {journal} {\bibinfo
  {journal} {Phys. Rev. A}\ }\textbf {\bibinfo {volume} {88}},\ \bibinfo
  {pages} {011803(R)} (\bibinfo {year} {2013}{\natexlab{a}})}\BibitemShut
  {NoStop}%
\bibitem [{\citenamefont {Grigoriev}\ \emph
  {et~al.}(2013{\natexlab{b}})\citenamefont {Grigoriev}, \citenamefont
  {Varault}, \citenamefont {Boudarham}, \citenamefont {Stout}, \citenamefont
  {Wenger},\ and\ \citenamefont {Bonod}}]{grigoriev2013singular}%
  \BibitemOpen
  \bibfield  {author} {\bibinfo {author} {\bibfnamefont {V.}~\bibnamefont
  {Grigoriev}}, \bibinfo {author} {\bibfnamefont {S.}~\bibnamefont {Varault}},
  \bibinfo {author} {\bibfnamefont {G.}~\bibnamefont {Boudarham}}, \bibinfo
  {author} {\bibfnamefont {B.}~\bibnamefont {Stout}}, \bibinfo {author}
  {\bibfnamefont {J.}~\bibnamefont {Wenger}}, \ and\ \bibinfo {author}
  {\bibfnamefont {N.}~\bibnamefont {Bonod}},\ }\href@noop {} {\bibfield
  {journal} {\bibinfo  {journal} {Phys. Rev. A}\ }\textbf {\bibinfo {volume}
  {88}},\ \bibinfo {pages} {063805} (\bibinfo {year}
  {2013}{\natexlab{b}})}\BibitemShut {NoStop}%
\bibitem [{\citenamefont {Krasnok}\ \emph {et~al.}(2019)\citenamefont
  {Krasnok}, \citenamefont {Baranov}, \citenamefont {Li}, \citenamefont {Miri},
  \citenamefont {Monticone},\ and\ \citenamefont {Alu}}]{Krasnok2019}%
  \BibitemOpen
  \bibfield  {author} {\bibinfo {author} {\bibfnamefont {A.}~\bibnamefont
  {Krasnok}}, \bibinfo {author} {\bibfnamefont {D.}~\bibnamefont {Baranov}},
  \bibinfo {author} {\bibfnamefont {H.}~\bibnamefont {Li}}, \bibinfo {author}
  {\bibfnamefont {M.-A.}\ \bibnamefont {Miri}}, \bibinfo {author}
  {\bibfnamefont {F.}~\bibnamefont {Monticone}}, \ and\ \bibinfo {author}
  {\bibfnamefont {A.}~\bibnamefont {Alu}},\ }\href@noop {} {\bibfield
  {journal} {\bibinfo  {journal} {Adv. Opt. Photonics}\ }\textbf {\bibinfo
  {volume} {11}},\ \bibinfo {pages} {892} (\bibinfo {year} {2019})}\BibitemShut
  {NoStop}%
\bibitem [{\citenamefont {Pichler}\ \emph {et~al.}(2019)\citenamefont
  {Pichler}, \citenamefont {K{\"u}hmayer}, \citenamefont {Bohm}, \citenamefont
  {Brandstotter}, \citenamefont {Ambichl}, \citenamefont {Kuhl},\ and\
  \citenamefont {Rotter}}]{Pichler2019}%
  \BibitemOpen
  \bibfield  {author} {\bibinfo {author} {\bibfnamefont {K.}~\bibnamefont
  {Pichler}}, \bibinfo {author} {\bibfnamefont {M.}~\bibnamefont {K{\"u}hmayer}},
  \bibinfo {author} {\bibfnamefont {J.}~\bibnamefont {Bohm}}, \bibinfo
  {author} {\bibfnamefont {A.}~\bibnamefont {Brandstotter}}, \bibinfo {author}
  {\bibfnamefont {P.}~\bibnamefont {Ambichl}}, \bibinfo {author} {\bibfnamefont
  {U.}~\bibnamefont {Kuhl}}, \ and\ \bibinfo {author} {\bibfnamefont
  {S.}~\bibnamefont {Rotter}},\ }\href@noop {} {\bibfield  {journal} {\bibinfo
  {journal} {Nature}\ }\textbf {\bibinfo {volume} {567}},\ \bibinfo {pages}
  {351} (\bibinfo {year} {2019})}\BibitemShut {NoStop}%
\bibitem [{\citenamefont {Sweeney}\ \emph {et~al.}(2020)\citenamefont
  {Sweeney}, \citenamefont {Hsu},\ and\ \citenamefont {Stone}}]{Sweeney2020}%
  \BibitemOpen
  \bibfield  {author} {\bibinfo {author} {\bibfnamefont {W.~R.}\ \bibnamefont
  {Sweeney}}, \bibinfo {author} {\bibfnamefont {C.~W.}\ \bibnamefont {Hsu}}, \
  and\ \bibinfo {author} {\bibfnamefont {A.~D.}\ \bibnamefont {Stone}},\
  }\href@noop {} {\bibfield  {journal} {\bibinfo  {journal} {Phys. Rev. A}\
  }\textbf {\bibinfo {volume} {102}},\ \bibinfo {pages} {063511} (\bibinfo
  {year} {2020})}\BibitemShut {NoStop}%
\bibitem [{\citenamefont {F.~Imani}\ \emph {et~al.}(2020)\citenamefont
  {F.~Imani}, \citenamefont {Smith},\ and\ \citenamefont {del
  Hougne}}]{imani2020perfect}%
  \BibitemOpen
  \bibfield  {author} {\bibinfo {author} {\bibfnamefont {M.}~\bibnamefont
  {F.~Imani}}, \bibinfo {author} {\bibfnamefont {D.~R.}\ \bibnamefont {Smith}},
  \ and\ \bibinfo {author} {\bibfnamefont {P.}~\bibnamefont {del Hougne}},\
  }\href@noop {} {\bibfield  {journal} {\bibinfo  {journal} {Adv. Funct.
  Mater.}\ }\textbf {\bibinfo {volume} {30}},\ \bibinfo {pages} {2005310}
  (\bibinfo {year} {2020})}\BibitemShut {NoStop}%
\bibitem [{\citenamefont {Frazier}\ \emph {et~al.}(2020)\citenamefont
  {Frazier}, \citenamefont {Antonsen~Jr}, \citenamefont {Anlage},\ and\
  \citenamefont {Ott}}]{frazier2020wavefront}%
  \BibitemOpen
  \bibfield  {author} {\bibinfo {author} {\bibfnamefont {B.~W.}\ \bibnamefont
  {Frazier}}, \bibinfo {author} {\bibfnamefont {T.~M.}\ \bibnamefont
  {Antonsen~Jr}}, \bibinfo {author} {\bibfnamefont {S.~M.}\ \bibnamefont
  {Anlage}}, \ and\ \bibinfo {author} {\bibfnamefont {E.}~\bibnamefont {Ott}},\
  }\href@noop {} {\bibfield  {journal} {\bibinfo  {journal} {Phys. Rev.
  Research}\ }\textbf {\bibinfo {volume} {2}},\ \bibinfo {pages} {043422}
  (\bibinfo {year} {2020})}\BibitemShut {NoStop}%
\bibitem [{\citenamefont {del Hougne}\ \emph
  {et~al.}(2021{\natexlab{b}})\citenamefont {del Hougne}, \citenamefont {Yeo},
  \citenamefont {Besnier},\ and\ \citenamefont {Davy}}]{delHougne2020CPA}%
  \BibitemOpen
  \bibfield  {author} {\bibinfo {author} {\bibfnamefont {P.}~\bibnamefont {del
  Hougne}}, \bibinfo {author} {\bibfnamefont {K.~B.}\ \bibnamefont {Yeo}},
  \bibinfo {author} {\bibfnamefont {P.}~\bibnamefont {Besnier}}, \ and\
  \bibinfo {author} {\bibfnamefont {M.}~\bibnamefont {Davy}},\ }\href@noop {}
  {\bibfield  {journal} {\bibinfo  {journal} {Laser Photonics Rev.}\ }\textbf
  {\bibinfo {volume} {15}},\ \bibinfo {pages} {2000471} (\bibinfo {year}
  {2021}{\natexlab{b}})}\BibitemShut {NoStop}%
\bibitem [{\citenamefont {del Hougne}\ \emph
  {et~al.}(2021{\natexlab{c}})\citenamefont {del Hougne}, \citenamefont {Yeo},
  \citenamefont {Besnier},\ and\ \citenamefont {Davy}}]{del2021coherent}%
  \BibitemOpen
  \bibfield  {author} {\bibinfo {author} {\bibfnamefont {P.}~\bibnamefont {del
  Hougne}}, \bibinfo {author} {\bibfnamefont {K.~B.}\ \bibnamefont {Yeo}},
  \bibinfo {author} {\bibfnamefont {P.}~\bibnamefont {Besnier}}, \ and\
  \bibinfo {author} {\bibfnamefont {M.}~\bibnamefont {Davy}},\ }\href@noop {}
  {\bibfield  {journal} {\bibinfo  {journal} {Phys. Rev. Lett.}\ }\textbf
  {\bibinfo {volume} {126}},\ \bibinfo {pages} {193903} (\bibinfo {year}
  {2021}{\natexlab{c}})}\BibitemShut {NoStop}%
\bibitem [{\citenamefont {Sol}\ \emph {et~al.}(2022)\citenamefont {Sol},
  \citenamefont {Smith},\ and\ \citenamefont {del Hougne}}]{sol2021meta}%
  \BibitemOpen
  \bibfield  {author} {\bibinfo {author} {\bibfnamefont {J.}~\bibnamefont
  {Sol}}, \bibinfo {author} {\bibfnamefont {D.~R.}\ \bibnamefont {Smith}}, \
  and\ \bibinfo {author} {\bibfnamefont {P.}~\bibnamefont {del Hougne}},\
  }\href@noop {} {\bibfield  {journal} {\bibinfo  {journal} {Nat. Commun.}\
  }\textbf {\bibinfo {volume} {13}},\ \bibinfo {pages} {1713} (\bibinfo {year}
  {2022})}\BibitemShut {NoStop}%
\bibitem [{\citenamefont {Sol}\ \emph {et~al.}(2023)\citenamefont {Sol},
  \citenamefont {Alhulaymi}, \citenamefont {Stone},\ and\ \citenamefont {del
  Hougne}}]{sol2022reflectionless}%
  \BibitemOpen
  \bibfield  {author} {\bibinfo {author} {\bibfnamefont {J.}~\bibnamefont
  {Sol}}, \bibinfo {author} {\bibfnamefont {A.}~\bibnamefont {Alhulaymi}},
  \bibinfo {author} {\bibfnamefont {A.~D.}\ \bibnamefont {Stone}}, \ and\
  \bibinfo {author} {\bibfnamefont {P.}~\bibnamefont {del Hougne}},\
  }\href@noop {} {\bibfield  {journal} {\bibinfo  {journal} {Sci. Adv.}\
  }\textbf {\bibinfo {volume} {9}},\ \bibinfo {pages} {eadf0323} (\bibinfo
  {year} {2023})}\BibitemShut {NoStop}%
\bibitem [{\citenamefont {Delage}\ \emph {et~al.}(2022)\citenamefont {Delage},
  \citenamefont {Pascal}, \citenamefont {Sokoloff},\ and\ \citenamefont
  {Mazi{\`e}res}}]{delage2022experimental}%
  \BibitemOpen
  \bibfield  {author} {\bibinfo {author} {\bibfnamefont {T.}~\bibnamefont
  {Delage}}, \bibinfo {author} {\bibfnamefont {O.}~\bibnamefont {Pascal}},
  \bibinfo {author} {\bibfnamefont {J.}~\bibnamefont {Sokoloff}}, \ and\
  \bibinfo {author} {\bibfnamefont {V.}~\bibnamefont {Mazi{\`e}res}},\
  }\href@noop {} {\bibfield  {journal} {\bibinfo  {journal} {J. Appl. Phys.}\
  }\textbf {\bibinfo {volume} {132}},\ \bibinfo {pages} {153105} (\bibinfo
  {year} {2022})}\BibitemShut {NoStop}%
\bibitem [{\citenamefont {Delage}\ \emph {et~al.}(2023)\citenamefont {Delage},
  \citenamefont {Sokoloff}, \citenamefont {Pascal}, \citenamefont
  {Mazi{\`e}res}, \citenamefont {Krasnok},\ and\ \citenamefont
  {Callegari}}]{delage2023reflectionless}%
  \BibitemOpen
  \bibfield  {author} {\bibinfo {author} {\bibfnamefont {T.}~\bibnamefont
  {Delage}}, \bibinfo {author} {\bibfnamefont {J.}~\bibnamefont {Sokoloff}},
  \bibinfo {author} {\bibfnamefont {O.}~\bibnamefont {Pascal}}, \bibinfo
  {author} {\bibfnamefont {V.}~\bibnamefont {Mazi{\`e}res}}, \bibinfo {author}
  {\bibfnamefont {A.}~\bibnamefont {Krasnok}}, \ and\ \bibinfo {author}
  {\bibfnamefont {T.}~\bibnamefont {Callegari}},\ }\href@noop {} {\bibfield
  {journal} {\bibinfo  {journal} {arXiv preprint arXiv:2306.03071}\ } (\bibinfo
  {year} {2023})}\BibitemShut {NoStop}%
\bibitem [{\citenamefont {Chen}\ and\ \citenamefont
  {Anlage}(2022)}]{chen2022use}%
  \BibitemOpen
  \bibfield  {author} {\bibinfo {author} {\bibfnamefont {L.}~\bibnamefont
  {Chen}}\ and\ \bibinfo {author} {\bibfnamefont {S.~M.}\ \bibnamefont
  {Anlage}},\ }\href@noop {} {\bibfield  {journal} {\bibinfo  {journal} {Phys.
  Rev. E}\ }\textbf {\bibinfo {volume} {105}},\ \bibinfo {pages} {054210}
  (\bibinfo {year} {2022})}\BibitemShut {NoStop}%
\bibitem [{\citenamefont {Mandelshtam}\ and\ \citenamefont
  {Taylor}(1997)}]{Mandelshtam1997}%
  \BibitemOpen
  \bibfield  {author} {\bibinfo {author} {\bibfnamefont {V.~A.}\ \bibnamefont
  {Mandelshtam}}\ and\ \bibinfo {author} {\bibfnamefont {H.~S.}\ \bibnamefont
  {Taylor}},\ }\href@noop {} {\bibfield  {journal} {\bibinfo  {journal} {J.
  Chem. Phys.}\ }\textbf {\bibinfo {volume} {107}},\ \bibinfo {pages} {6756}
  (\bibinfo {year} {1997})}\BibitemShut {NoStop}%
\bibitem [{\citenamefont {Kuhl}\ \emph {et~al.}(2008)\citenamefont {Kuhl},
  \citenamefont {Hohmann}, \citenamefont {Main},\ and\ \citenamefont
  {Stockmann}}]{Kuhl2008}%
  \BibitemOpen
  \bibfield  {author} {\bibinfo {author} {\bibfnamefont {U.}~\bibnamefont
  {Kuhl}}, \bibinfo {author} {\bibfnamefont {R.}~\bibnamefont {Hohmann}},
  \bibinfo {author} {\bibfnamefont {J.}~\bibnamefont {Main}}, \ and\ \bibinfo
  {author} {\bibfnamefont {H.~J.}\ \bibnamefont {Stockmann}},\ }\href@noop {}
  {\bibfield  {journal} {\bibinfo  {journal} {Phys. Rev. Lett.}\ }\textbf
  {\bibinfo {volume} {100}},\ \bibinfo {pages} {254101} (\bibinfo {year}
  {2008})}\BibitemShut {NoStop}%
\bibitem [{\citenamefont {Davy}\ and\ \citenamefont {Genack}(2018)}]{Davy2018}%
  \BibitemOpen
  \bibfield  {author} {\bibinfo {author} {\bibfnamefont {M.}~\bibnamefont
  {Davy}}\ and\ \bibinfo {author} {\bibfnamefont {A.~Z.}\ \bibnamefont
  {Genack}},\ }\href@noop {} {\bibfield  {journal} {\bibinfo  {journal} {Nat.
  Comm.}\ }\textbf {\bibinfo {volume} {9}},\ \bibinfo {pages} {4714} (\bibinfo
  {year} {2018})}\BibitemShut {NoStop}%
\bibitem [{\citenamefont {Weyl}(1911)}]{Weyl1911}%
  \BibitemOpen
  \bibfield  {author} {\bibinfo {author} {\bibfnamefont {H.}~\bibnamefont
  {Weyl}},\ }\href@noop {} {\bibfield  {journal} {\bibinfo  {journal}
  {{Nachrichten von der Gesellschaft der Wissenschaften zu G\"ottingen,
  Mathematisch-Physikalische Klasse}}\ }\textbf {\bibinfo {volume} {1911}},\
  \bibinfo {pages} {110} (\bibinfo {year} {1911})}\BibitemShut {NoStop}%
\bibitem [{\citenamefont {Arendt}\ \emph {et~al.}(2009)\citenamefont {Arendt},
  \citenamefont {Nittka}, \citenamefont {Peter}, \citenamefont {Steiner},\ and\
  \citenamefont {Schleich}}]{Arendt}%
  \BibitemOpen
  \bibfield  {author} {\bibinfo {author} {\bibfnamefont {W.}~\bibnamefont
  {Arendt}}, \bibinfo {author} {\bibfnamefont {R.}~\bibnamefont {Nittka}},
  \bibinfo {author} {\bibfnamefont {W.}~\bibnamefont {Peter}}, \bibinfo
  {author} {\bibfnamefont {F.}~\bibnamefont {Steiner}}, \ and\ \bibinfo
  {author} {\bibfnamefont {W.}~\bibnamefont {Schleich}},\ }\href@noop {} {\emph
  {\bibinfo {title} {{Mathematical Analysis of Evolution, Information, and
  Complexity, Weyl's Law}}}}\ (\bibinfo  {publisher} {Wiley-VCH, Weinheim,
  Germany},\ \bibinfo {year} {2009})\ pp.\ \bibinfo {pages} {1--71}\BibitemShut
  {NoStop}%
\bibitem [{\citenamefont {Pierrat}\ \emph {et~al.}(2014)\citenamefont
  {Pierrat}, \citenamefont {Ambichl}, \citenamefont {Gigan}, \citenamefont
  {Haber}, \citenamefont {Carminati},\ and\ \citenamefont
  {Rotter}}]{Pierrat2014}%
  \BibitemOpen
  \bibfield  {author} {\bibinfo {author} {\bibfnamefont {R.}~\bibnamefont
  {Pierrat}}, \bibinfo {author} {\bibfnamefont {P.}~\bibnamefont {Ambichl}},
  \bibinfo {author} {\bibfnamefont {S.}~\bibnamefont {Gigan}}, \bibinfo
  {author} {\bibfnamefont {A.}~\bibnamefont {Haber}}, \bibinfo {author}
  {\bibfnamefont {R.}~\bibnamefont {Carminati}}, \ and\ \bibinfo {author}
  {\bibfnamefont {S.}~\bibnamefont {Rotter}},\ }\href@noop {} {\bibfield
  {journal} {\bibinfo  {journal} {Proc. Natl. Acad. Sci.}\ }\textbf {\bibinfo
  {volume} {111}},\ \bibinfo {pages} {17765} (\bibinfo {year}
  {2014})}\BibitemShut {NoStop}%
\bibitem [{\citenamefont {Wigner}(1955)}]{Wigner1955}%
  \BibitemOpen
  \bibfield  {author} {\bibinfo {author} {\bibfnamefont {E.~P.}\ \bibnamefont
  {Wigner}},\ }\href@noop {} {\bibfield  {journal} {\bibinfo  {journal} {Phys.
  Rev.}\ }\textbf {\bibinfo {volume} {98}},\ \bibinfo {pages} {145} (\bibinfo
  {year} {1955})}\BibitemShut {NoStop}%
\bibitem [{\citenamefont {Smith}(1960)}]{Smith1960}%
  \BibitemOpen
  \bibfield  {author} {\bibinfo {author} {\bibfnamefont {F.~T.}\ \bibnamefont
  {Smith}},\ }\href@noop {} {\bibfield  {journal} {\bibinfo  {journal} {Phys.
  Rev.}\ }\textbf {\bibinfo {volume} {118}},\ \bibinfo {pages} {349} (\bibinfo
  {year} {1960})}\BibitemShut {NoStop}%
\bibitem [{\citenamefont {Fan}\ and\ \citenamefont
  {Kahn}(2005)}]{fan2005principal}%
  \BibitemOpen
  \bibfield  {author} {\bibinfo {author} {\bibfnamefont {S.}~\bibnamefont
  {Fan}}\ and\ \bibinfo {author} {\bibfnamefont {J.~M.}\ \bibnamefont {Kahn}},\
  }\href@noop {} {\bibfield  {journal} {\bibinfo  {journal} {Opt. Lett.}\
  }\textbf {\bibinfo {volume} {30}},\ \bibinfo {pages} {135} (\bibinfo {year}
  {2005})}\BibitemShut {NoStop}%
\bibitem [{\citenamefont {Rotter}\ \emph {et~al.}(2011)\citenamefont {Rotter},
  \citenamefont {Ambichl},\ and\ \citenamefont {Libisch}}]{Rotter2011}%
  \BibitemOpen
  \bibfield  {author} {\bibinfo {author} {\bibfnamefont {S.}~\bibnamefont
  {Rotter}}, \bibinfo {author} {\bibfnamefont {P.}~\bibnamefont {Ambichl}}, \
  and\ \bibinfo {author} {\bibfnamefont {F.}~\bibnamefont {Libisch}},\
  }\href@noop {} {\bibfield  {journal} {\bibinfo  {journal} {Phys. Rev. Lett.}\
  }\textbf {\bibinfo {volume} {106}},\ \bibinfo {pages} {120602} (\bibinfo
  {year} {2011})}\BibitemShut {NoStop}%
\bibitem [{\citenamefont {Davy}\ \emph {et~al.}(2015)\citenamefont {Davy},
  \citenamefont {Shi}, \citenamefont {Wang}, \citenamefont {Cheng},\ and\
  \citenamefont {Genack}}]{Davy2015}%
  \BibitemOpen
  \bibfield  {author} {\bibinfo {author} {\bibfnamefont {M.}~\bibnamefont
  {Davy}}, \bibinfo {author} {\bibfnamefont {Z.}~\bibnamefont {Shi}}, \bibinfo
  {author} {\bibfnamefont {J.}~\bibnamefont {Wang}}, \bibinfo {author}
  {\bibfnamefont {X.}~\bibnamefont {Cheng}}, \ and\ \bibinfo {author}
  {\bibfnamefont {A.~Z.}\ \bibnamefont {Genack}},\ }\href@noop {} {\bibfield
  {journal} {\bibinfo  {journal} {Phys. Rev. Lett.}\ }\textbf {\bibinfo
  {volume} {114}},\ \bibinfo {pages} {033901} (\bibinfo {year}
  {2015})}\BibitemShut {NoStop}%
\bibitem [{\citenamefont {B\"ohm}\ \emph {et~al.}(2018)\citenamefont {B\"ohm},
  \citenamefont {Brandst\"otter}, \citenamefont {Ambichl}, \citenamefont
  {Rotter},\ and\ \citenamefont {Kuhl}}]{Boehm2018}%
  \BibitemOpen
  \bibfield  {author} {\bibinfo {author} {\bibfnamefont {J.}~\bibnamefont
  {B\"ohm}}, \bibinfo {author} {\bibfnamefont {A.}~\bibnamefont
  {Brandst\"otter}}, \bibinfo {author} {\bibfnamefont {P.}~\bibnamefont
  {Ambichl}}, \bibinfo {author} {\bibfnamefont {S.}~\bibnamefont {Rotter}}, \
  and\ \bibinfo {author} {\bibfnamefont {U.}~\bibnamefont {Kuhl}},\ }\href@noop
  {} {\bibfield  {journal} {\bibinfo  {journal} {Phys. Rev. A}\ }\textbf
  {\bibinfo {volume} {97}},\ \bibinfo {pages} {021801(R)} (\bibinfo {year}
  {2018})}\BibitemShut {NoStop}%
\bibitem [{\citenamefont {Huang}\ \emph {et~al.}(2022)\citenamefont {Huang},
  \citenamefont {Kang},\ and\ \citenamefont {Genack}}]{huang2022wave}%
  \BibitemOpen
  \bibfield  {author} {\bibinfo {author} {\bibfnamefont {Y.}~\bibnamefont
  {Huang}}, \bibinfo {author} {\bibfnamefont {Y.}~\bibnamefont {Kang}}, \ and\
  \bibinfo {author} {\bibfnamefont {A.~Z.}\ \bibnamefont {Genack}},\
  }\href@noop {} {\bibfield  {journal} {\bibinfo  {journal} {Phys. Rev. Res.}\
  }\textbf {\bibinfo {volume} {4}},\ \bibinfo {pages} {013102} (\bibinfo {year}
  {2022})}\BibitemShut {NoStop}%
\bibitem [{\citenamefont {Asano}\ \emph {et~al.}(2016)\citenamefont {Asano},
  \citenamefont {Bliokh}, \citenamefont {Bliokh}, \citenamefont {Kofman},
  \citenamefont {Ikuta}, \citenamefont {Yamamoto}, \citenamefont {Kivshar},
  \citenamefont {Yang}, \citenamefont {Imoto}, \citenamefont {{\"O}zdemir},\ and\
  \citenamefont {Nori}}]{Asano2016}%
  \BibitemOpen
  \bibfield  {author} {\bibinfo {author} {\bibfnamefont {M.}~\bibnamefont
  {Asano}}, \bibinfo {author} {\bibfnamefont {K.~Y.}\ \bibnamefont {Bliokh}},
  \bibinfo {author} {\bibfnamefont {Y.~P.}\ \bibnamefont {Bliokh}}, \bibinfo
  {author} {\bibfnamefont {A.~G.}\ \bibnamefont {Kofman}}, \bibinfo {author}
  {\bibfnamefont {R.}~\bibnamefont {Ikuta}}, \bibinfo {author} {\bibfnamefont
  {T.}~\bibnamefont {Yamamoto}}, \bibinfo {author} {\bibfnamefont {Y.~S.}\
  \bibnamefont {Kivshar}}, \bibinfo {author} {\bibfnamefont {L.}~\bibnamefont
  {Yang}}, \bibinfo {author} {\bibfnamefont {N.}~\bibnamefont {Imoto}},
  \bibinfo {author} {\bibfnamefont {{\c{S}}.~K.}\ \bibnamefont {{\"O}zdemir}}, \ and\
  \bibinfo {author} {\bibfnamefont {F.}~\bibnamefont {Nori}},\ }\href@noop {}
  {\bibfield  {journal} {\bibinfo  {journal} {Nat. Commun.}\ }\textbf {\bibinfo
  {volume} {7}},\ \bibinfo {pages} {13488} (\bibinfo {year}
  {2016})}\BibitemShut {NoStop}%
\bibitem [{\citenamefont {Chen}\ \emph
  {et~al.}(2021{\natexlab{a}})\citenamefont {Chen}, \citenamefont {Anlage},\
  and\ \citenamefont {Fyodorov}}]{ChenPRL2021}%
  \BibitemOpen
  \bibfield  {author} {\bibinfo {author} {\bibfnamefont {L.}~\bibnamefont
  {Chen}}, \bibinfo {author} {\bibfnamefont {S.~M.}\ \bibnamefont {Anlage}}, \
  and\ \bibinfo {author} {\bibfnamefont {Y.~V.}\ \bibnamefont {Fyodorov}},\
  }\href@noop {} {\bibfield  {journal} {\bibinfo  {journal} {Phys. Rev. Lett.}\
  }\textbf {\bibinfo {volume} {127}},\ \bibinfo {pages} {204101} (\bibinfo
  {year} {2021}{\natexlab{a}})}\BibitemShut {NoStop}%
\bibitem [{\citenamefont {Chen}\ \emph
  {et~al.}(2021{\natexlab{b}})\citenamefont {Chen}, \citenamefont {Anlage},\
  and\ \citenamefont {Fyodorov}}]{ChenPRE2021}%
  \BibitemOpen
  \bibfield  {author} {\bibinfo {author} {\bibfnamefont {L.}~\bibnamefont
  {Chen}}, \bibinfo {author} {\bibfnamefont {S.~M.}\ \bibnamefont {Anlage}}, \
  and\ \bibinfo {author} {\bibfnamefont {Y.~V.}\ \bibnamefont {Fyodorov}},\
  }\href@noop {} {\bibfield  {journal} {\bibinfo  {journal} {Phys. Rev. E}\
  }\textbf {\bibinfo {volume} {103}},\ \bibinfo {pages} {L050203} (\bibinfo
  {year} {2021}{\natexlab{b}})}\BibitemShut {NoStop}%
\bibitem [{\citenamefont {Davy}\ \emph {et~al.}(2021)\citenamefont {Davy},
  \citenamefont {K{\"u}hmayer}, \citenamefont {Gigan},\ and\ \citenamefont
  {Rotter}}]{Davy2021mean}%
  \BibitemOpen
  \bibfield  {author} {\bibinfo {author} {\bibfnamefont {M.}~\bibnamefont
  {Davy}}, \bibinfo {author} {\bibfnamefont {M.}~\bibnamefont {K{\"u}hmayer}},
  \bibinfo {author} {\bibfnamefont {S.}~\bibnamefont {Gigan}}, \ and\ \bibinfo
  {author} {\bibfnamefont {S.}~\bibnamefont {Rotter}},\ }\href@noop {}
  {\bibfield  {journal} {\bibinfo  {journal} {Commun. Phys.}\ }\textbf
  {\bibinfo {volume} {4}},\ \bibinfo {pages} {85} (\bibinfo {year}
  {2021})}\BibitemShut {NoStop}%
\end{thebibliography}

\begin{thebibliography}{1}%
\makeatletter
\providecommand \@ifxundefined [1]{%
 \@ifx{#1\undefined}
}%
\providecommand \@ifnum [1]{%
 \ifnum #1\expandafter \@firstoftwo
 \else \expandafter \@secondoftwo
 \fi
}%
\providecommand \@ifx [1]{%
 \ifx #1\expandafter \@firstoftwo
 \else \expandafter \@secondoftwo
 \fi
}%
\providecommand \natexlab [1]{#1}%
\providecommand \enquote  [1]{``#1''}%
\providecommand \bibnamefont  [1]{#1}%
\providecommand \bibfnamefont [1]{#1}%
\providecommand \citenamefont [1]{#1}%
\providecommand \href@noop [0]{\@secondoftwo}%
\providecommand \href [0]{\begingroup \@sanitize@url \@href}%
\providecommand \@href[1]{\@@startlink{#1}\@@href}%
\providecommand \@@href[1]{\endgroup#1\@@endlink}%
\providecommand \@sanitize@url [0]{\catcode `\\12\catcode `\$12\catcode
  `\&12\catcode `\#12\catcode `\^12\catcode `\_12\catcode `\%12\relax}%
\providecommand \@@startlink[1]{}%
\providecommand \@@endlink[0]{}%
\providecommand \url  [0]{\begingroup\@sanitize@url \@url }%
\providecommand \@url [1]{\endgroup\@href {#1}{\urlprefix }}%
\providecommand \urlprefix  [0]{URL }%
\providecommand \Eprint [0]{\href }%
\providecommand \doibase [0]{http://dx.doi.org/}%
\providecommand \selectlanguage [0]{\@gobble}%
\providecommand \bibinfo  [0]{\@secondoftwo}%
\providecommand \bibfield  [0]{\@secondoftwo}%
\providecommand \translation [1]{[#1]}%
\providecommand \BibitemOpen [0]{}%
\providecommand \bibitemStop [0]{}%
\providecommand \bibitemNoStop [0]{.\EOS\space}%
\providecommand \EOS [0]{\spacefactor3000\relax}%
\providecommand \BibitemShut  [1]{\csname bibitem#1\endcsname}%
\let\auto@bib@innerbib\@empty
\bibitem [{\citenamefont {Chong}\ \emph {et~al.}(2010)\citenamefont {Chong},
  \citenamefont {Ge}, \citenamefont {Cao},\ and\ \citenamefont
  {Stone}}]{chong2010coherent}%
  \BibitemOpen
  \bibfield  {author} {\bibinfo {author} {\bibfnamefont {Y.}~\bibnamefont
  {Chong}}, \bibinfo {author} {\bibfnamefont {L.}~\bibnamefont {Ge}}, \bibinfo
  {author} {\bibfnamefont {H.}~\bibnamefont {Cao}}, \ and\ \bibinfo {author}
  {\bibfnamefont {A.~D.}\ \bibnamefont {Stone}},\ }\href@noop {} {\bibfield
  {journal} {\bibinfo  {journal} {Phys. Rev. Lett.}\ }\textbf {\bibinfo
  {volume} {105}},\ \bibinfo {pages} {053901} (\bibinfo {year}
  {2010})}\BibitemShut {NoStop}%
\end{thebibliography}

\providecommand{\noopsort}[1]{}\providecommand{\singleletter}[1]{#1}%

\end{document}